\documentclass[hyper]{JHEP}
\usepackage{amssymb,bm,latexsym,amsmath}

\title{D1 string dynamics in curved backgrounds with fluxes}
\author{Aritra Banerjee, Sagar Biswas* and  Rashmi R. Nayak \\
Department of Physics,\\Indian Institute of Technology Kharagpur,\\
Kharagpur-721 302, India\\ * Department of Physics, R. K. M. Vidyamandira,\\ Belur Math, Howrah 711 202,India\\
Email: \email{aritra@phy.iitkgp.ernet.in ; biswas.sagar09iitkgp@gmail.com ; rashmi.string@gmail.com
}}

\abstract{We study various rotating and oscillating D-string
configurations in some general backgrounds with fluxes. In
particular, we look for solutions to the equations of motion of
various rigidly rotating D-strings in $AdS_3$ background with
mixed flux, and in the intersecting D-brane geometries. We find
out relations among various conserved charges corresponding to the
breathing and rotating D-string configurations.}


\begin{document}
\section{Introduction and summary}
The $AdS/CFT$ correspondence \cite{Maldacena:1997re,Witten:1998qj, Gubser:1998bc},
relates string states in $AdS$ spaces to some conformal
field theory (CFT) living on its boundary. In general, it is very difficult to
find the full spectrum of states in the string side and then compare it with the anomalous
dimensions of the operators on the CFT side. This has been tough  even for the very well studied example of the
$\mathcal{N}= 4$ supersymmetric Yang-Mills (SYM) theory in four dimensions and dual type
IIB superstring in the compactified $AdS_5$ space. One of the direct ways to check the
correspondence beyond the supergravity approximation has been to study various classical string
solutions in varieties of target space geometries and using the dispersion relation of such
strings in the large charge limit, one could look for boundary operators dual to them
\footnote{For a review of semiclassical strings in $AdS/CFT$ one may look at
\cite{Tseytlin:2004xa}.}. This has been one of the main ways to establish the $AdS/CFT$
dictionary in many cases. Instead of probe classical fundamental strings, there have also
been many studies on various exact string backgrounds using probe Dp-branes. This provides
a novel way of understanding string theory in curved background as the Dp-brane couples to RR
fluxes present.

 One of the first attempts in this direction was to study D-branes in the $SL(2,\mathbb{R})$
 Wess-Zumino-Witten (WZW) model and its discreet orbifolds \cite{Balog:1988jb}. The corresponding
 target space geometries are mainly $AdS_3$ with non-trivial NS-NS fluxes and
 3d $AdS$ black holes \cite{ Banados:1992wn, Banados:1992gq}. This was extensively studied
 in a lot of works and the WZW D-branes are now well understood from both CFT boundary states
 and the target space viewpoint  \cite{klim,as,fffs,sta,ars,bds,paw,ars2} . These brane worldvolumes
 are also shown to carry non-trivial gauge fields. Also coupling to background NS-NS fluxes resists
 a D-string probe from reducing to a point particle \cite{bds,paw}. However, it was also shown that under
 a supercritical worldvolume electric field, a circular oscillating D-string can never reach the boundary
 of $AdS$ space \cite{Bachas:2000fr}. In a related development, D1 strings rotating in $\mathbb{R}\times S^2$ and
 $\mathbb{R}\times S^3$ were studied in detail by using Dirac-Born-Infeld (DBI) action \cite{Kluson:2007fr}
 to find giant magnon \cite{Hofman:2006xt} and single spike \cite{Ishizeki:2007we} like solutions
 on the D1-string. Giant magnons arise as rotating solutions in the string side corresponding to low lying
 spin-chain excitations in the dual field theory. Similarly single spike configurations are particular string
 solutions which correspond to a particular class of single trace operators in the field theory with large
 number of derivatives. Both of these appear generally as fundamental string solutions and  the D1-string
 analog in the presence of worldvolume gauge field indeed appears to have yet-unknown novel interpretation.

In general, D1 string solutions in Dp-brane backgrounds are rare due to the complicated nature of the
background and presence of dilaton and other RR fields. But this problem appears to be interesting in conjunction to
the prediction of \cite{Stepanchuk:2012xi} that a general Dp-brane background is non-integrable for extended
objects. However various well behaved probe brane solutions have been constructed in these backgrounds too. One example
is in \cite{Kluson:2006wa,Kluson:2007st}, where the system consisting of two stacks of the fivebranes in type IIB
theory that intersect on $\mathbb{R}^{1,1}$ (an Intersecting brane or I-brane \cite{Itzhaki:2005tu} ) has been
investigated using probe D-strings. The much discussed enhancement of symmetry in the near horizon geometry of such
systems have been shown to have profound impact on the D1-string worldvolume itself.

Motivated by the above studies, we move on to discuss probe D1 string solutions in a few general settings
with coupling to background fluxes. The first study we undertake is a natural generalization of the WZW D-brane
solutions. Recently, it was shown that the string theory on $AdS_3\times S^3$ supported by both NS-NS and RR
three-form fluxes is integrable \cite{Cagnazzo:2012se, Wulff:2014kja}. There has been proposals of S-matrix on
this background  \cite{Hoare:2013pma,Hoare:2013ida, Bianchi:2014rfa,Hoare:2013lja,Babichenko:2014yaa,Borsato:2014hja,
Lloyd:2014bsa} and various classical string solutions \cite{Hoare:2013lja,David:2014qta,Ahn:2014tua,misc9,Hernandez:2014eta,
Hernandez:2015nba,Banerjee:2015bia,Banerjee:2015qeq} have been constructed. The NS-NS flux in this  background is
parameterized by a number $q$ with $0 \leq q \leq 1$, while the RR 3-form is proportional to $\hat{q}=\sqrt{1-q^2}$.
As the $q$ interpolates between 0 to 1, the solution interpolates between that of a pure RR background and a pure
NS-NS backgound described by the usual WZW model. However, for intermediate values of $q$, the exact description
of the string sigma model is not known. In any case the study of open string integrability in this background
is still lacking, most probably due to dearth of D-brane boundary conditions. However, recently the integrability
of a D1-string on the group manifold with mixed three form fluxes has been argued  and Lax connections have been
constructed \cite{Kluson:2015lia} with the condition that dilaton and Ramond-Ramond zero forms are constants. 
It is worth noting that
the analysis has only been performed at the bosonic level and it remains to show whether the full action is integrable.
Still, even in the bosonic sector, it remains an interesting problem to analyse dynamics of D1-strings in this background. We
will probe this backgound with a bound state of oscillating D1 strings and F-strings with non-trivial gauge field
on the D1 worldvolume. One of the apparent outcomes of the analysis based on the DBI action is that the periodically expanding and
contracting $(1,n)$ string has a possibility of reaching the boundary of $AdS_3$ in finite time in contrast to
the probe D-string motion in the WZW model with only NS-NS fluxes.

The later parts of the paper is devoted to study of rotating D1 string solutions in various Dp brane backgrounds.
In particular we will talk about the near horizon geometry of a stack of D5 branes and two stacks of D5 branes
that intersect on a line. We solve the equations of motion for D1 string keeping couplings with dilatons and WZ
terms in mind. It is shown that in proper limits, the combination of properly regularised conserved quantities
actually give rise to giant magnon and single spike like dispersion relations as in the case of fundamental strings.
We speculate that these exact solutions correspond to S-dual fundamental string  solutions in NS5 brane background
\cite{Biswas:2011wu} and in NS5-NS5' brane intersections \cite{Biswas:2012wu}.

The rest of the paper is organised as follows. In section 2 we will discuss the motion of a $(m,n)$ string in the mixed-flux
$AdS_3$ background. In section 3, we will move on to the motion of a rotating D1-string in the near horizon geometry
of a stack of D5 branes. We will show that the conserved charges give rise to
giant magnon or single spike-like dispersion relations in two different limits. We will discuss a similar configuration
in the intersecting D5-D5' brane background in section 4. The equations of motion appear to be very complicated in that case,
but with certain simplifications we will be again able to find giant magnon or single spike-like dispersion relations
for the properly regularised version of the charges. We conclude with some outlook in section 5.

\section{Circular $(m,n)$ strings on $AdS_3$ with mixed 3-form fluxes }
The mixed flux background is a solution of the type IIB action with a $AdS_3\times S^3\times T^4$ geometry, although
the compact manifold will not be interesting here. The solution has both RR and NS-NS fluxes along the $AdS$ and $S$
directions. In this section we will put the $S^3$ coordinates to be constant and consider motion only along the $AdS_3$.
The background and the relevant fluxes are as follows,
\begin{eqnarray}
    ds^2 &=& -\cosh^2\rho dt^2 + d\rho^2 + \sinh^2\rho d\phi^2 \ , \nonumber \\ B_{(2)} &=& q\cosh^2\rho~ dt \wedge d\phi \ ,
    \>\>\> C_{(2)} = \sqrt{1-q^2} \cosh^2\rho ~dt \wedge d\phi \ .
\end{eqnarray}
The dilaton $\Phi$ is constant and can be set to zero. Also, the 
$AdS$ radius has been properly chosen here. The above
background can easily be shown to be a solution of the type IIB
field equations. \footnote{Note that there is a gauge freedom in
choosing the two-form B-field since the supergravity equations of
motion involves $H_{(3)}= dB_{(2)}$ only. For example, in the case of a three sphere the NS-NS flux is
defined upto an additive constant in the form $-\frac{q}{2}(\cos
2\theta +c)$. This constant can be fixed using physical considerations. We have fixed the constant here
for the $AdS_3$ case so that the fields have the mentioned form. For details
one could see the arguments presented in \cite{Hoare:2013lja}.}

Now, we want to discuss the motion of a bound state of $m$ D1-strings and $n$ F-strings in this background as a general string state
in type IIB theory would readily be the bound state of the two. For simplicity we will choose $m=1$ and argue that the analysis can be
extended to other more general string state too. The DBI action for a D1-brane can be written as,
\begin{equation}
    S = -T_D \int d^2\xi e^{-\phi} \sqrt{ - \det (\hat{g} + \hat{B} + 2\pi \alpha^{\prime}F)} +
    \frac{T_D}{2} \int d^2\xi \epsilon^{\alpha\beta}C_{\alpha\beta} \ .
\end{equation}
where $\hat{g}$ and $\hat{B}$ are the pullback of the background metric and the NS-NS fluxes and $F$ is the field strength for the
worldvolume gauge field. Here $T_D$ is the D1-string
tension $T_D= 1/(2\pi \alpha' g_s) $, with $g_s$ being the string coupling. In contrast, remember that the F-string tension
was simply $T_F= 1/2\pi \alpha' $. So in weak coupling regime we can write $T_D>> T_F$. Also, $\epsilon^{\alpha \beta}$
is the usual antisymmetric tensor with $\epsilon^{0 1} = 1$. Since the mixed flux background has the
constant (or zero) dilaton, the Lagrangian density takes the form,
\begin{equation}
    \mathcal{L} = -T_D \Big[\sqrt{ - \det (\hat{g} + \hat{B} + 2\pi \alpha^{\prime}F)} -
    \frac{1}{2} \epsilon^{\alpha\beta}C_{\alpha\beta} \Big] \ .
\end{equation}
We choose the following ansatz for a `breathing' mode of the string:
\begin{eqnarray}
    t = \xi^0 \ , \>\>\> \rho (\xi^0, \xi^1) = \rho(\xi^0) \ , \>\>\> \phi = \xi^1 \ .
\end{eqnarray}
The ansatz is like the one for a simple circular F-string motion and can be easliy shown to be consistent with
the equations of motion here. Using the above ansatz we can show,
\begin{eqnarray}
    -\det(\hat{g} + \hat{B} + 2\pi \alpha^{\prime}F) &=& \sinh^2\rho (\cosh^2\rho - (\partial_0\rho)^2) -
    (q\cosh^2\rho - 2\pi \alpha^{\prime} \partial_0 A_{\phi})^2 \nonumber \\ &=& -\det \> \hat {g} - \mathcal{F}_{\phi t}^2 \ ,
\end{eqnarray}
where we have
\begin{equation}
 \det \> \hat{g} = -\sinh^2\rho (\cosh^2\rho - (\partial_0\rho)^2)
\end{equation}
and
\begin{equation}
\mathcal{F}_{\phi t}
= q\cosh^2\rho - 2\pi\alpha^{\prime} \partial_0 A_{\phi}.
\end{equation}
Now, we can rewrite the lagrangian density as,
\begin{equation}
    \mathcal{L} = -T_D [\sqrt{-\det \> \hat{g} - \mathcal{F}^2_{\phi t}} - \sqrt{1 - q^2} \cosh^2\rho] \ .
\end{equation}
Conjugate momentum of the Wilson line $A_{\phi}$ is a quantized constant of the motion given by,
\begin{equation}
    \frac{1}{2\pi} \Pi_{\phi} = \frac{\partial \mathcal{L}}{\partial(\partial_0A_{\phi})} =
    \frac{-2\pi\alpha^{\prime} T_D \mathcal{F}_{\phi t}}{\sqrt{-\det \hat {g} - \mathcal{F}_{\phi t}^2}} = -n \ . \label{wilson}
\end{equation}
The integer $n$ is the number of oriented fundamental strings bound to the D-string.
The effective tension of $(m,n)$ circular string is given by,
\begin{equation}
    T_{(m,n)} = \sqrt{m^2T^2_D + n^2T^2_F} \ .
\end{equation}
From the above expression and (\ref{wilson}) we can write,
\begin{equation}
 T_{(1,n)}= T_D \sqrt{1+ \frac{(2\pi\alpha^{\prime})^2 T_F^2 \mathcal{F}_{\phi t}^2}{(-\det \hat{g}-\mathcal{F}_{\phi t}^2)}}.
\end{equation}
Combining all expression we get the following relation which we will later use to simplify other expressions,
\begin{equation}
    \frac{T_{(1,n)}}{\sqrt{-\det\> \hat{g}}} = \frac{T_D}{\sqrt{-\det\> \hat{g} -
    \mathcal{F}^2_{\phi t}}} = \frac{n}{2\pi\alpha^{\prime} \mathcal{F}_{\phi t}} \ .
\end{equation}
The other constant of the motion is the energy, as measured by an observer
sitting at the center of $AdS_3$,
\begin{eqnarray}
   E &=& 2\pi \Big( \frac{\partial \mathcal{L}}{\partial(\partial_0\rho)} \partial_0\rho +
   \frac{\partial \mathcal{L}}{\partial(\partial_0 A_{\phi})} \partial_0 A_{\phi} - \mathcal{L} \Big)
   \nonumber \\ &=& 2\pi T_D \Big[ \frac{\sinh^2\rho \cosh^2\rho - q\cosh^2\rho(q\cosh^2\rho - 2\pi \alpha^{\prime}
   \partial_0 A_{\phi})}{\sqrt{-\det \> \hat{g} - \mathcal{F}^2_{\phi t}}} - \sqrt{1 - q^2}\cosh^2\rho\Big] \ . \nonumber \\
\end{eqnarray}
Putting the energy in a more suggestive form we can get,
\begin{equation}
    E = \frac{2\pi T_{(1,n)}\sinh\rho \cosh^2\rho}{\sqrt{\cosh^2\rho - (\partial_0\rho)^2}} -
    2\pi \cosh^2\rho (nqT_F + \sqrt{1-q^2}T_D) \ . \label{energy}
\end{equation}
Above equation exhibits the  competing terms of the potential energy: the
blue-shifted mass, and the interaction with the B-field and C-field potential. Using the above we can write the
equation of motion for $\rho$  as,
\begin{equation}
\partial_0 \rho = \frac{\cosh\rho}{E+ C_2\cosh^2\rho}\left[ (E+ C_2\cosh^2\rho)^2-
C_1^2 \sinh^2\rho \cosh^2\rho \right]^{1/2} \label{eomrho}
\end{equation}
Where the constants
\begin{equation}
C_1= 2\pi T_{(1,n)}~~~~~
C_2= 2\pi (nqT_F + \sqrt{1-q^2}T_D)
\end{equation}
Note that $C_1>C_2$ is still valid as we have $q \in [0,1]$.
Now our interest is to find a large energy solution, i.e. a limit where the string becomes `long' and tries to reach the
boundary of $AdS$ in finite time. In this limit we can solve the radial equation perturbatively in orders of
$\frac{1}{E}$. Expanding \ref{eomrho}, we get,
\begin{equation}
\partial_0 \rho = \cosh \rho - \frac{C_1^2 \sinh ^2\rho  \cosh ^3\rho }{2 E ^2} +
\frac{C_1^2 C_2 \sinh ^2\rho  \cosh ^5\rho }{E^3} + \mathcal{O}\left(\frac{1}{E^4}\right)
\end{equation}
To solve this order by order we take the expression for $\rho$
\begin{equation}
\rho = \rho^{(0)} + \frac{\rho^{(1)}}{E^2} + \frac{\rho^{(2)}}{E^3} + \mathcal{O}\left(\frac{1}{E^4}\right)
\end{equation}
Putting the above back in \ref{eomrho} and expanding we get the following set of coupled differential equations for
the different orders
\begin{eqnarray}
\partial_0 {\rho}^{(0)}&=& \cosh\rho^{(0)}, \nonumber\\
\partial_0 {\rho}^{(1)}&=& \rho^{(1)}\sinh\rho^{(0)}- \frac{1}{2}C_1^2\cosh^3\rho^{(0)}\sinh^2\rho^{(0)}, \nonumber\\
\partial_0 {\rho}^{(2)} &=& \rho^{(2)}\sinh\rho^{(0)}+ C_1^2 C_2\cosh^5\rho^{(0)}\sinh^2\rho^{(0)} \ . \nonumber\\
\end{eqnarray}
We solve the first equation to find
\begin{equation}
\rho^{(0)} = \sinh^{-1}\tan\tau,
\end{equation}
which is a periodically expanding and contracting solution where the string goes out to a maximum radius.
We use this solution iteratively in the other two equations and find the total solution using the boundary condition
$\rho(0) = 0$ to write the perturbative solution
\begin{equation}
\rho(\tau) =  \sinh^{-1}\tan\tau -\frac{1}{6 E^2}C_1^2\sec\tau \tan^3\tau + \frac{1}{15 E^3}C_1^2 C_2(4 + \cos 2\tau)
\tan^3\tau \sec^3\tau + \mathcal{O}\left(\frac{1}{E^4}\right)
\end{equation}

 Now the dynamics is incredibly difficult as there are many parameters involved like $T_F$, $T_D$, $n$ and $q$.
 The dynamics varies widely for different values of these parameters.
 With higher order corrections (which are suppressed in the large energy limit), the $(1,n)$ string goes through quasi-periodic
 motion.

 Now to find the maximum radius of the circular D-string we note that the extremum would occur at $\partial_0\rho  = 0$, which
 leads us to the following,
 \begin{equation}
  E+ C_2\cosh^2\rho_m - C_1 \sinh\rho_m \cosh\rho_m = 0.
 \end{equation}
With a little algebra we can find that,
\begin{equation}
 \rho_m = \frac{1}{2}~\ln \left[\frac{2E+ C_2 \pm \sqrt{4 E^2 + 2 E C_2 + C_1^2}}{(C_1-C_2)}\right].
\end{equation}
To prove that the above corresponds to the maximum value of $\rho$ we can explicitly show that
\begin{equation}
 \partial_0^2\rho(\rho_m) < 0.
\end{equation}
Since we are interested in the large $E$ limit of the solution, we can write the maximum radius approximately in a
simpler form by putting in the expressions for $C_{1,2}$,
\begin{equation}
\rho_m  \simeq \frac{1}{2}~\ln \left[\frac{2 E}{\pi(T_{(1,n)}- n q T_F- \sqrt{1- q^2}T_D)} \right] \label{rhomax}
\end{equation}
If one keeps all other parameters fixed and increases $q$ from $0$, it can easily be seen that the maximum value of $\rho$
actually increases. So, we can say that with increasing NS-NS flux the large energy string actually gets `fatter'.
However, we are interested to know whether the string becomes a `long' one.
Now interestingly enough, one can see that the expression for maximum radius diverges when
\begin{equation}
 T_{(1,n)}\rightarrow n q T_F+ \sqrt{1- q^2}T_D. \label{limit}
\end{equation}
Note that diverging $\rho_m$ means that the strings can actually expand upto the boundary of the $AdS_3$.
For the case of pure
NS-NS flux i.e. $q=1$ as discussed in \cite{Bachas:2000fr}, the $\rho_m$ can only diverge if $ T_{(1,n)}\rightarrow n  T_F$, so that
the string becomes a purely fundamental `long' string. It was made clear in \cite{Bachas:2000fr} that a general $(m,n)$ string can never
reach the boundary of $AdS$ as the first term in (\ref{energy}) diverges faster near the boundary than the NS-NS flux potential
term for $q=1$, making the energy effectively infinite. It is quite clear that for pure NS-NS case the two terms in (\ref{energy})
would cancel in the asymptotic region to produce a finite contribution, only if the string is purely fundamental. However here
we also have contribution from  the RR term, so the string might not need to be purely fundamental to acquire a finite energy near
the boundary. In our example, keeping in mind that $ T_{(1,n)}= \sqrt{T_D^2 + n^2 T_F^2}$,
we can easily see the following,
\begin{eqnarray}
 T_D + n T_F &>& T_{(1,n)}, \nonumber\\
\text{and}~~ T_D + n T_F &>& n q T_F+ \sqrt{1- q^2}T_D ;~~~~~ 0<q<1.
\end{eqnarray}
So in any case the condition (\ref{limit}) actually could be physical in a particular region of the parameter space
and the boundary might not remain forbidden region for the $(1,n)$ string when it couples to both NS-NS and RR fluxes.
In fact if we impose an equality in the (\ref{limit}), it simply yields the condition
\begin{equation}
 n\sqrt{1-q^2}~T_F= \pm q T_D,
\end{equation}
for the maximum radius of the string to diverge. We shall not be bothering about the negative sign here. It is very clear that in the 
weak coupling region (small $g_s$), we can transform the above condition into,
\begin{equation}
 n\sqrt{\frac{1-q^2}{q^2}}>>1.
\end{equation}
As we have $n$ to be a positive finite integer, the above condition can only be satisfied if the value of $q$
is small. As $q\to 1$, the above inequality
is violated since the equality associated with (\ref{limit}) loses meaning for this case and can
only be true if the string becomes a purely fundamental one. This supports the discussion we presented earlier.
This is a unique observation for such strings in this mixed flux background and have to be investigated into details
using other approaches. 

\section{Rotating D1-string on D5-branes}
Let us first start with the discussion of generic Dp brane backgrounds. For an arbitrary $p$, we can write
the general supergravity solution as follows,
\begin{eqnarray}
  ds^2 &=& h^{-\frac{1}{2}}(\vec{x})(-dt^2 + d\vec{y}~^2)+ h^{\frac{1}{2}}(\vec{x})~d\vec{x}~^2, \nonumber\\
  e^{2\phi}&=& h(\vec{x})^{\frac{3-p}{2}}, \nonumber\\
  C^{p+1}&=& \pm \left( \frac{1}{h(\vec{x})}-1\right) ~dt \wedge dy^1 \wedge...\wedge dy^p, \nonumber\\
  F^{p+2}&=&dC^{p+1}= \mp h^{-2}\partial_j h~dx^j \wedge dt \wedge dy^1 \wedge...\wedge dy^p, \nonumber\\
  \tilde{F}^{8-p}&=& \pm \partial_j h~{i_{\hat{x}^j}}(dx^1 \wedge...\wedge dx^{9-p}),\nonumber\\
  h(\vec{x})&=& 1+ \frac{\mu}{7-p}\left(\frac{l_s}{r}\right)^{7-p}. \label{dbrane}
\end{eqnarray}
Let us remind ourselves the various expressions used in the above equation. Here $p$ spatial coordinates 
$y^k$ run parallel to the worldvolume while the transverse space is given by $(9-p)$ coordinates labelled
by $x^i$. The specified harmonic function $ h(\vec{x})$ is valid for $p= 0,1,2...,6$ with 
$r^2 = \sum_{i=1}^{(9-p)}x^i x_i$ being the radial distance along the transverse space. The dilaton is 
given by $\phi$ and the RR form is given by $C^{p+1}$. The RR field
strength is given by $F^{p+2}$ while the hodge dual field strength is given by $\tilde{F}^{8-p}$. The 
action of $i_{\hat{x}^j}$ on the volume form of the transverse space signifies a inner product 
with a unit vector pointing in the $x^j$ direction.

For a D1-string moving in a general Dp-brane background, the DBI action is given by,
\begin{equation}
    S=-T_1\int d\xi^0d\xi^1e^{-\phi}\sqrt{-\det A_{\alpha \beta}} + T_1\int d\xi^0d\xi^1 C_{01}
\end{equation}
where $A_{\alpha \beta}=g_{\alpha\beta}=\partial_{\alpha}X^M\partial_{\beta}X^N g_{MN}$
is the induced metric on the worldvolume and $\alpha, \beta = \xi^0, \xi^1$. We here put
the worldvolume gauge field to be zero for simplicity. $C_{01}$ is the two form RR potential
 coupled to the D1 string.

The background in which we are interested in is a stack of N D5-branes, which is given by the following metric
, dilaton and RR six-form:
\begin{eqnarray}
    ds^2 &=& h^{-\frac{1}{2}}\Big[-dt^2 + \sum_{i=1}^{5}dx_i^2\Big] + h^{\frac{1}{2}} \Big[dr^2 + r^2(d\theta^2 +
    \sin^2\theta d\phi_1^2 + \cos^2\theta d\phi_2^2)\Big] \nonumber \\ e^{\phi} &=& h^{-\frac{1}{2}}, \>\>\> C_{012345}
    = \frac{k}{k+r^2}, \>\>\> h=1+\frac{k}{r^2}, \>\>\> k=g_sNl_s^2.
\end{eqnarray}
Here $l_s$ is the string length scale and $N$ is the number of branes in the stack.
In the near horizon limit $r \to 0$, so in the harmonic function, the second term dominates 
and it becomes $h = 1 + \frac{k}{r^2} \approx \frac{k}{r^2}$
and also the six form potential $C_{012345} = 1 + \mathcal{O}(r^2)$.
We now rescale the coordinates in the following way
\begin{equation}
t \to \sqrt{k}t ~~~ \and~~~ x_i \to \sqrt{k}x_i,
\end{equation}
and following this, the metric and fields reduce to the form,
\begin{eqnarray}
    ds^2 &=& \sqrt{k}r\Big[-dt^2 + \sum_{i=1}^{5}dx_i^2 + \frac{dr^2}{r^2} + d\theta^2 +
    \sin^2\theta d\phi_1^2 + \cos^2\theta d\phi_2^2 \Big] \nonumber \\ e^{\phi} &=& \frac{r}{\sqrt{k}},
    \>\>\> C_{012345} = 1, \>\>\> k=g_sNl_s^2.
\end{eqnarray}
This six form background RR field does not couple to a D1-string. However, the dual of this six form i.e.
the `magnetic' two form RR field $(\tilde{C}_{\phi_1\phi_2})$  will couple to the of D1-string
as WZ contribution\footnote{For detailed discussion on this point one could see the discussion in  \cite{Burgess:2003mm}.}.
For our convenience we define $\rho = \ln r$, and transform the the metric and the dilaton to,
\begin{eqnarray}
    ds^2 &=& \sqrt{k} e^{\rho}\Big[-dt^2 + \sum_{i=1}^{5}dx_i^2 + d\rho^2 + d\theta^2 + \sin^2\theta d\phi_1^2 +
    \cos^2\theta d\phi_2^2 \Big] \nonumber \\ e^{\phi} &=& \frac{e^{\rho}}{\sqrt{k}}, \>\>\> \tilde{C}_{\phi_1\phi_2}
    = 2k\sin^2\theta, \>\>\> k=g_sNl_s^2.
\end{eqnarray}
Now, it is to be noted that the RR two form $(\tilde{C}_{\phi_1\phi_2})$ has been written upto an additive
constant as we can calculate the field strength $F^{(3)}$ from the (\ref{dbrane}). This 3-form turns out simply to be
the volume form of the transverse sphere itself. Now for our probe D1 string, 
the induced worldvolume metric components are given by,
\begin{eqnarray}
    A_{00} &=& \sqrt{k}e^{\rho}\Big[-(\partial_0t)^2 + \sum (\partial_0x_i)^2 + (\partial_0 \rho)^2 + (\partial_0\theta)^2
    + \sin^2\theta (\partial_0\phi_1)^2 + \cos^2\theta (\partial_0\phi_2)^2 \Big] \nonumber \\
 A_{11} &=& \sqrt{k}e^{\rho}\Big[-(\partial_1t)^2 + \sum (\partial_1x_i)^2 + (\partial_1 \rho)^2 + (\partial_1\theta)^2
 + \sin^2\theta (\partial_1\phi_1)^2 + \cos^2\theta (\partial_1\phi_2)^2 \Big] \nonumber \\
 A_{01} &=& A_{10} = \sqrt{k}e^{\rho}\Big[-(\partial_0t)(\partial_1t) + \sum (\partial_0x_i)(\partial_1x_i) +
 (\partial_0 \rho)(\partial_1 \rho) + (\partial_0\theta)(\partial_1\theta) \nonumber \\ &+& \sin^2\theta (\partial_0\phi_1)
 (\partial_1\phi_1) + \cos^2\theta (\partial_0\phi_2)(\partial_1\phi_2) \Big] \ .
\end{eqnarray}
Therefore we can write the total Lagrangian-density as,
\begin{eqnarray}
    \mathcal{L} &=& -   T_1e^{-\phi}\sqrt{-\det A_{\alpha \beta}} + \frac{T_1}{2} \epsilon^{\alpha\beta}
    \partial_{\alpha}X^{M}\partial_{\beta}X^{N} C_{MN}\nonumber \\ &=& -T_1k\Big[[-(\partial_0t)(\partial_1t)
    + \sum (\partial_0x_i)(\partial_1x_i) + (\partial_0 \rho)(\partial_1 \rho) + (\partial_0\theta)(\partial_1\theta)
    \nonumber \\ &+& \sin^2\theta (\partial_0\phi_1)(\partial_1\phi_1) + \cos^2\theta (\partial_0\phi_2)(\partial_1\phi_2)]^2
    -[-(\partial_0t)^2 + \sum (\partial_0x_i)^2 \nonumber \\ &+& (\partial_0 \rho)^2 + (\partial_0\theta)^2 +
    \sin^2\theta (\partial_0\phi_1)^2 + \cos^2\theta (\partial_0\phi_2)^2][-(\partial_1t)^2 \nonumber \\ &+&
    \sum (\partial_1x_i)^2 + (\partial_1 \rho)^2 + (\partial_1\theta)^2 + \sin^2\theta (\partial_1\phi_1)^2 +
    \cos^2\theta (\partial_1\phi_2)^2]\Big]^{\frac{1}{2}} \nonumber \\ &+& 2T_1k\sin^2\theta(\partial_0\phi_1\partial_1\phi_2
    - \partial_0\phi_2\partial_1\phi_1) \ .
\end{eqnarray}
Before solving the Euler-Lagrange equations,
\begin{equation}
    \partial_0\Big(\frac{\partial \mathcal{L}}{\partial(\partial_0 X)}\Big) +
    \partial_1\Big(\frac{\partial \mathcal{L}}{\partial(\partial_1 X)}\Big) = \frac{\partial \mathcal{L}}{\partial X} \ ,
\end{equation}
we choose a rotating ansatz for the probe brane,
\begin{eqnarray}
    t &=& \kappa \xi^0, \>\>\> x_i = \nu_i \xi^0, \>\>\> i=1,2,3,4,5, \>\>\> \rho = m\xi^0, \nonumber \\
    \theta &=& \theta(\xi^1), \>\>\> \phi_1 = \omega_1\xi^0 + \xi^1, \>\>\> \phi_2 = \omega_2\xi^0 + \phi_2(\xi^1) \ .
    \label{ansatzD}
\end{eqnarray}
The above ansatz can be thought of as the generalized version of the rigidly rotating string parameterisation described 
in, for example, \cite{Ishizeki:2007we}. 
Solving the equations of motion for $\phi_1$ and $\phi_2$ and eliminating $\mathcal{L}$ from the respective equations we get,

\begin{eqnarray}\label{ephi}
    \frac{\partial \phi_2}{\partial \xi^1} = \frac{\sin^2\theta [(c_2-2\omega_2\sin^2\theta)\omega_1\omega_2\cos^2\theta +
    (c_3+2\omega_1\sin^2\theta)\omega_2^2\cos^2\theta -\alpha^2(c_3+2\omega_1\sin^2\theta)]}{\cos^2\theta [(c_3+2\omega_1\sin^2\theta)
    \omega_1\omega_2\sin^2\theta + (c_2-2\omega_2\sin^2\theta)\omega_1^2\sin^2\theta -\alpha^2(c_2-2\omega_2\sin^2\theta)]} \ . \nonumber \\  \label{eqn 003}
\end{eqnarray}
Again solving for $t$, we get the equation for $\theta$,
\begin{eqnarray}\label{etheta}
     \Big(\frac{\partial \theta}{\partial \xi^1}\Big)^2 &=& \frac{(c_1^2-\kappa^2)(\omega_1\sin^2\theta +
     \omega_2 \partial_1\phi_2\cos^2\theta)^2}{c_1^2\{-\alpha^2  + \omega_1^2\sin^2\theta + \omega_2^2\cos^2\theta\}}
     \nonumber \\ &-&  \{ \sin^2\theta + (\partial_1\phi_2)^2\cos^2\theta\} \ . \label{eqn 004}
\end{eqnarray}
where $c_1$, $c_2$ and $c_3$ are carefully chosen integration constants and $\alpha^2 = \kappa^2 - m^2 - \sum_{i=1}^5\nu_i\nu^i$.

Certainly, the equation of motion for $\theta$ is quite complicated and difficult to solve in general due to the 
presence of all kinds of constants.
 In what follows we will discuss two limiting cases from the above set of equations corresponding to single spike
 and giant magnon solutions. In \cite{Ishizeki:2007we} it was shown that the giant magnon and single spike strings can
 be thought of two limits of the same system of equations. We will show that this is the same for our case also.
 
 Let us outline the procedure to impose conditions on the equations of motion in a simple schematic way. The main problem
 lies in the behaviour of $\frac{\partial \phi_2}{\partial \xi^1}$ at $\theta = \frac{\pi}{2}$ as there is a $\cos^2\theta$
 sitting in the denominator. To talk about the behaviour of $\frac{\partial \theta}{\partial \xi^1}$ in this limit, we will
 then define
 \begin{equation}
  \left(\frac{\partial \phi_2}{\partial \xi^1}\right)^2 \cos^2\theta \bigg{|}_{\theta = \frac{\pi}{2}}
  = \mathcal{P}^2.
 \end{equation}
Now we want to have particular conditions imposed on $\frac{\partial \theta}{\partial \xi^1}$ too. We can see that
\begin{equation}
 \frac{\partial \theta}{\partial \xi^1}
\bigg{|}_{\theta = \frac{\pi}{2}}
= \frac{c_1^2\alpha^2 - \kappa^2\omega_1^2 + \mathcal{P}^2 c_1^2(\alpha^2- \omega_1^2)}{c_1^2(-\alpha^2+ \omega_1^2)}.
\end{equation}
There are two very interesting limits that can be considered here. The first one is to demand that
$\frac{\partial \theta}{\partial \xi^1} \to 0$ as $\theta \to \frac{\pi}{2}$, which in turn demands that
$c_1\alpha = \kappa \omega_1$. Also it is required in this case that $\mathcal{P}^2\to 0$ i.e.
$\frac{\partial \phi_2}{\partial \xi^1}$
has to be regular, which in turn demands that $c_3 = -2\omega_1$. So, this set of relations between the constants will
implement the conditions mentioned above. The other limit may be taken as  
$\frac{\partial \theta}{\partial \xi^1} \rightarrow \infty$ as $\theta \to \frac{\pi}{2}$. It also demands that 
$\frac{\partial \phi_2}{\partial \xi^1} \to \infty$ in the same limit of $\theta$. Both of these 
can be realised by the set of relations $\alpha = \omega_1$ and 
$c_3 = -2\omega_1$. We will discuss these particular two conditions while finding out the constants of motion and the
relevant dispersion relations among them. We will see that the equations of motion considerably simplify when we put
in the relations between the constants.

\subsection{Single spike solution}
 In this section we will follow the procedure outlined in \cite{Ishizeki:2007we} and consider 
 constants of motions appropriately, demanding the following condition is satisfied,
 \begin{equation}
 \frac{\partial \theta}{\partial \xi^1} \rightarrow 0 ~~~
 \text{as} ~~
 \theta \to \frac{\pi}{2}.
 \end{equation}
 As we discussed earlier, this will also require $\frac{\partial \phi_2}{\partial \xi^1}$ to be regular.
 So combining these, we  get the relations between the constants $c_1\alpha = \kappa \omega_1$ 
 and $c_3 = -2\omega_1$. Note that the imposed conditions determine
 the value of integration constants $c_1$ and $c_3$ in terms of other parameters, but it doesn't fix the value of $c_2$. For the
 time being we can keep $c_2$ as it is. Later, we will try to fix a value of $c_2$ using other considerations.
 Using these the equations (\ref{ephi}) and (\ref{etheta}) transform to ,
\begin{equation}
    \frac{\partial \phi_2}{\partial \xi^1} = \frac{\omega_1(2\omega_2^2 - 2\alpha^2 - c_2\omega_2 )\sin^2\theta }
    {c_2\alpha^2 +(2\omega_1^2\omega_2 - 2\alpha^2\omega_2 - c_2\omega_1^2)\sin^2\theta} \ ,
\end{equation}
and
\begin{equation}
     \frac{\partial \theta}{\partial \xi^1} = \frac{\alpha \sqrt{(\omega_2^2 - \omega_1^2)\{4\alpha^2 -
     (c_2-2\omega_2)^2\}}\sin\theta\cos\theta \sqrt{\sin^2\theta - \sin^2\theta_0}}{c_2\alpha^2 +(2\omega_1^2\omega_2
     - 2\alpha^2\omega_2 - c_2\omega_1^2)\sin^2\theta} \ . \label{theta11}
\end{equation}
Where we have the upper limit on $\theta$ in the form,
\begin{equation}
\sin\theta_0 = \frac{c_2\alpha}{\sqrt{(\omega_2^2 - \omega_1^2)\{4\alpha^2 - (c_2-2\omega_2)^2\}}}.
\end{equation}
Now we can explicitly solve for $\theta$ equation from (\ref{theta11})to get the profile of the string, which
gives
\begin{eqnarray}
 \alpha \sqrt{(\omega_2^2 - \omega_1^2)\{4\alpha^2 -
     (c_2-2\omega_2)^2\}}~\xi^1 &=& \frac{c_2(\alpha^2-\omega_1^2) +(2\omega_1^2\omega_2
     - 2\alpha^2\omega_2) }{\cos\theta_0}\cosh^{-1}\left[\frac{\cos\theta_0}{\cos\theta}\right] \nonumber\\
     &-& \frac{c_2\alpha^2}{\sin\theta_0}\cos^{-1}\left[\frac{\sin\theta_0}{\sin\theta}\right].
\end{eqnarray}
From the above we can see $\theta= \theta_0$ can be thought of as the point exactly where the tip of the spike is situated. 
On  the other hand it is clear that at $\theta \to \frac{\pi}{2}$ we actually have $\xi_1 \to \infty$,
making this a valid single spike solution.
The largest spike goes up to the pole of the sphere where $\theta_0 = 0$.
Now we can calculate the conserved charges for the D-string motion using usual noether techniques,
\begin{eqnarray}
    E &=& -2\int_{\theta_0}^{\theta_1} \frac{\partial \mathcal{L}}{\partial_0 t} \frac{d\theta}{\partial_1\theta} =
    \frac{2T_1k\kappa (\alpha^2 - \omega_1^2) (c_2 - 2\omega_2)}{\alpha^2 \sqrt{(\omega_2^2 - \omega_1^2)\{4\alpha^2 -
    (c_2-2\omega_2)^2\}}} \int_{\theta_0}^{\frac{\pi}{2}} \frac{\sin\theta d\theta}{\cos\theta\sqrt{\sin^2\theta -
    \sin^2\theta_0}}, \nonumber \\ P_i &=& 2\int_{\theta_0}^{\theta_1} \frac{\partial \mathcal{L}}{\partial_0 x_i}
    \frac{d\theta}{\partial_1\theta} = \frac{2T_1k\nu_i (\alpha^2 - \omega_1^2) (c_2 - 2\omega_2)}{\alpha^2
    \sqrt{(\omega_2^2 - \omega_1^2)\{4\alpha^2 - (c_2-2\omega_2)^2\}}} \int_{\theta_0}^{\frac{\pi}{2}}
    \frac{\sin\theta d\theta}{\cos\theta\sqrt{\sin^2\theta - \sin^2\theta_0}}, \nonumber \\ D &=&
    2\int_{\theta_0}^{\theta_1} \frac{\partial \mathcal{L}}{\partial_0 \rho} \frac{d\theta}{\partial_1\theta}
    =\frac{2T_1k m (\alpha^2 - \omega_1^2) (c_2 - 2\omega_2)}{\alpha^2 \sqrt{(\omega_2^2 - \omega_1^2)\{4\alpha^2 -
    (c_2-2\omega_2)^2\}}} \int_{\theta_0}^{\frac{\pi}{2}} \frac{\sin\theta d\theta}{\cos\theta\sqrt{\sin^2\theta -
    \sin^2\theta_0}} \ . \nonumber \\
\end{eqnarray}
Here as usual, the $E$ corresponds to time translation invariance, and $P_i$'s correspond to shifts in $x_i$'s.
The other charge $D$ comes from translations in $\rho = \ln r$. In general this is not a symmetry of the
metric itself, but the way we have chosen the parameterisations in (\ref{ansatzD}) makes it clear that
these are indeed symmetries of the D1 string action.
All these quantities diverge due to the divergent integral. The angular momenta $J_1$ and $J_2$ that arise from the isometries
along $\phi_1$ and $\phi_2$ are given by,
\begin{eqnarray}
     J_1 &=& \frac{2T_1k\omega_1 [2(2\omega_2^2 - 2\alpha^2 - c_2\omega_2) - c_2\alpha]}{\alpha \sqrt{(\omega_2^2 -
     \omega_1^2)\{4\alpha^2 - (c_2-2\omega_2)^2\}}} \int_{\theta_0}^{\frac{\pi}{2}} \frac{\sin\theta \cos\theta d\theta}
     {\sqrt{\sin^2\theta - \sin^2\theta_0}} \nonumber \\ && - \frac{4T_1k\omega_1 [(2\omega_2^2 - 2\alpha^2 - c_2\omega_2)]}
     {\alpha \sqrt{(\omega_2^2 - \omega_1^2)\{4\alpha^2 - (c_2-2\omega_2)^2\}}}
\int_{\theta_0}^{\frac{\pi}{2}}\frac{\sin\theta d\theta}{\cos\theta \sqrt{\sin^2\theta - \sin^2\theta_0}} \ .
\end{eqnarray}
\begin{eqnarray}
     J_2 &=& \frac{2T_1k[\alpha(2\omega_1^2 - 2\omega_2^2 + c_2\omega_2) - 2(2\omega_1^2\omega_2 -
     2\alpha^2\omega_2 - c_2\omega_1^2)]}{\alpha \sqrt{(\omega_2^2 - \omega_1^2)\{4\alpha^2 -
     (c_2-2\omega_2)^2\}}} \int_{\theta_0}^{\frac{\pi}{2}} \frac{\sin\theta \cos\theta d\theta}
     {\sqrt{\sin^2\theta - \sin^2\theta_0}} \nonumber \\ && + \frac{2T_1k[2(c_2\alpha^2 + 2\omega_1^2\omega_2
     - 2\alpha^2\omega_2 - c_2\omega_1^2) ]}{\alpha \sqrt{(\omega_2^2 - \omega_1^2)\{4\alpha^2 -
     (c_2-2\omega_2)^2\}}} \int_{\theta_0}^{\frac{\pi}{2}}\frac{\sin\theta d\theta}{\cos\theta
     \sqrt{\sin^2\theta - \sin^2\theta_0}} \ .
\end{eqnarray}
The other quantity of interest is the angle deficit defined as
$\Delta\phi =2\int_{\theta_0}^{\theta_1}  \frac{d\theta}{\partial_1\theta}$ is given by,
\begin{eqnarray}
    \Delta\phi = \frac{2}{\alpha \sqrt{(\omega_2^2 - \omega_1^2)\{4\alpha^2 - (c_2 - 2\omega_2)^2\}}}
    && \Bigg[(c_2 - 2\omega_2)(\alpha^2 - \omega_1^2) \int_{\theta_0}^{\frac{\pi}{2}} \frac{\sin\theta d\theta}
    {\cos\theta \sqrt{\sin^2\theta - \sin^2\theta_0}} \nonumber \\ && + c_2\alpha^2 \int_{\theta_0}^{\frac{\pi}{2}}
    \frac{\cos\theta}{\sin\theta \sqrt{\sin^2\theta - \sin^2\theta_0}}\Bigg] \ ,
\end{eqnarray}
which is clearly divergent because of the first integral. However one can regularize the angle difference using
a combination of other conserved charges to remove the divergent part,
\begin{equation}
    (\Delta\phi)_{reg} = \Delta\phi - \frac{1}{T_1k} \sqrt{E^2 - D^2 - \sum_i P_i^2} = 2\cos^{-1}(\sin\theta_0) \ ,
\end{equation}
which implies
\begin{equation}
\sin\theta_0 = \cos (\frac{(\Delta\phi)_{reg}}{2}).
\end{equation}
We can also regularize both the angular momenta as,
\begin{eqnarray}
    (J_1)_{reg} &=& J_1 + \frac{2\omega_1(2\omega_2^2 - 2\alpha^2 - c_2\omega_2)}{(c_2 - 2\omega_2)(\alpha^2 -
    \omega_1^2)} \sqrt{E^2 - \sum P_i^2 - D^2} \nonumber \\ &=& \frac{2T_1k\omega_1(4\omega_2^2 - 4\alpha^2 -
    2c_2\omega_2 - c_2\alpha)}{\alpha \sqrt{(\omega_2^2 - \omega_1^2)\{4\alpha^2 - (c_2 - 2\omega_2)^2\}}} \cos\theta_0
\end{eqnarray}
and,
\begin{eqnarray}
    (J_2)_{reg} &=& J_2 - \frac{2(c_2\alpha^2 + 2\omega_1^2\omega_2 - 2\alpha^2\omega_2 - c_2\omega_1^2)}
    {(c_2 - 2\omega_2)(\alpha^2 - \omega_1^2)} \sqrt{E^2 - \sum P_i^2 - D^2} \nonumber \\ &=& \frac{2T_1k\{\alpha
    (2\omega_1^2 - 2\omega_2^2 + c_2\omega_2) -2(2\omega_1^2\omega_2 - 2\alpha^2\omega_2 - c_2\omega_1^2)\}}
    {\alpha \sqrt{(\omega_2^2 - \omega_1^2)\{4\alpha^2 - (c_2 - 2\omega_2)^2\}}} \cos\theta_0
\end{eqnarray}
This regularisation for charges of a single-spike D1-string can be compared to similar construction 
of F-strings in a background B field, considered in for example, \cite{Chen:2008vc}. It can actually be
shown that without the WZ term contribution, the angular momentum $J_1$ and $J_2$ are finite as can also
be seen in \cite{Ishizeki:2007we}. But with
finite background flux, they diverge and we have to regularise the charges to get a finite answer. 
These regularised angular momenta can be easily found to satisfy the dispersion relation,
\begin{equation}
    (J_2)_{reg} = \sqrt{(J_1)_{reg}^2 + f_1(\lambda)\sin^2\Big( \frac{(\Delta\phi)_{reg}}{2}\Big)} \ ,
\end{equation}
where $f_1(\lambda)$ is a complicated function of various constants and winding numbers and
 $N = \sqrt{\lambda}$ is the effective 't Hooft coupling.  However,
if we choose $c_2 = 2\omega_2 - \alpha$, then the complicated function reduces to $f_1(\lambda) = \frac{3\lambda}{\pi^2}$
and the dispersion relation looks like,
\begin{equation}
    (J_2)_{reg} = \sqrt{(J_1)_{reg}^2 + \frac{3\lambda}{\pi^2}\sin^2\Big( \frac{(\Delta\phi)_{reg}}{2}\Big)} \ ,
\end{equation}
which matches exactly with the dispersion relation obtained by studying fundamental rotating single-spike string solutions on
NS5-branes \cite{Biswas:2011wu}.

\subsection{Giant Magnon}
 In this case we demand the situation where both $\frac{\partial \theta}{\partial \xi^1}$ 
 and $\frac{\partial \phi_2}{\partial \xi^1}$ diverge.
 Using these conditions yield the relations between the constants $\alpha = \omega_1$ and $c_3 =
 -2\omega_1$. In this case also the integration constant $c_2$ remains undetermined. As in the previous case we will
 put the value of $c_2$ by hand to reduce the dispersion relation. Using these conditions we write down the equations
 in the form,
\begin{equation}
    \frac{\partial \phi_2}{\partial \xi^1} = \frac{(2\omega_2^2 - 2\omega_1^2 - c_2\omega_2)\sin^2\theta }
    {c_2\omega_1\cos^2\theta} \ ,
  \end{equation}
and
\begin{equation}
     \frac{\partial \theta}{\partial \xi^1} = \frac{\sin\theta \sqrt{\sin^2\theta - \sin^2\theta_1}}
     {\cos\theta \sin\theta_1} \ ,
\end{equation}
where the upper limit for $\theta$ is,
\begin{equation}
\sin\theta_1 = \frac{c_1 c_2\omega_1}{\sqrt{(\omega_2^2 - \omega_1^2)(4c_1^2\omega_1^2 - c_2^2\kappa^2 -
4\kappa^2\omega_2^2 + 4c_2\kappa^2\omega_2)}}.
\end{equation}
We can now easily integrate the $\theta$ equation to find the string profile as follows
\begin{equation}
 \xi^1 = \cos^{-1}\left[\frac{\sin\theta_1}{\sin\theta}\right],
\end{equation}
which gives a finite range of $\xi^1$ with $|\xi^1| \leq \frac{\pi}{2}-\theta_1$ as expected.

In this case we can construct the conserved charges in the usual sense,
\begin{eqnarray}
    E &=& \frac{2T_1k(\kappa^2-c_1^2)(c_2 - 2\omega_2)\sin\theta_1}{c_1c_2 \omega_1} \int^{\theta_1}_{\frac{\pi}{2}}
    \frac{\sin\theta d\theta}{\cos\theta\sqrt{\sin^2\theta - \sin^2\theta_1}}, \nonumber \\ P_i &=&
    \frac{2T_1k\nu_i(\kappa^2 - c_1^2)(c_2 - 2\omega_2)\sin\theta_1}{c_1c_2\kappa \omega_1}
    \int^{\theta_1}_{\frac{\pi}{2}} \frac{\sin\theta d\theta}{\cos\theta\sqrt{\sin^2\theta -
    \sin^2\theta_1}}, \nonumber \\ D &=&  \frac{2T_1km(\kappa^2 - c_1^2)(c_2 - 2\omega_2)\sin\theta_1}
    {c_1c_2\kappa \omega_1} \int^{\theta_1}_{\frac{\pi}{2}} \frac{\sin\theta d\theta}
    {\cos\theta\sqrt{\sin^2\theta - \sin^2\theta_1}} \ .
\end{eqnarray}
As in single-spike case before all these quantities diverge due to the divergent integral. Among the angular
momenta we can write,
\begin{eqnarray}
    J_1 &=& \frac{2T_1k\{\omega_1(\kappa^2 - c_1^2)(c_2 - 2\omega_2) + 2c_1\kappa (2\omega_2^2 - 2\omega_1^2
    - c_2\omega_2)\}}{c_1c_2\kappa \omega_1} \int_{\frac{\pi}{2}}^{\theta_1} \frac{\sin\theta_1\sin\theta
    d\theta}{\cos\theta \sqrt{\sin^2\theta - \sin^2\theta_1}} \nonumber \\ && - \frac{2T_1k\{2c_1\kappa
    (2\omega_2^2 - 2\omega_1^2 - c_2\omega_2) + \omega_1(2c_1^2\omega_2 + c_2\kappa^2 - 2\kappa^2\omega_2)\}}
    {c_1c_2\kappa \omega_1} \int_{\frac{\pi}{2}}^{\theta_1} \frac{\sin\theta_1\sin\theta\cos\theta d\theta}
    { \sqrt{\sin^2\theta - \sin^2\theta_1}}  \ , \nonumber \\
\end{eqnarray}
Now this $J_1$ is divergent, but on the other hand,
\begin{equation}
    J_2  = \frac{2T_1k[2c_1c_2\kappa\omega_1 - 2c_1^2\omega_1^2 - \kappa^2 \omega_2(c_2 - 2\omega_2)]}
    {c_1c_2\kappa \omega_1} \sin\theta_1\cos\theta_1 \ ,
\end{equation}
is finite. Also the angle deficit can be shown to be finite,
\begin{eqnarray}
    \Delta\phi = -2\cos^{-1}(\sin\theta_1) \ .
\end{eqnarray}
which implies $\sin\theta_1 = \cos (\frac{(\Delta\phi)}{2})$.
If we define the divergent quantity,
\begin{equation}
    \tilde{E} = \frac{\omega_1(\kappa^2 - c_1^2)(c_2 - 2\omega_2) + 2c_1\kappa(2\omega_2^2 - 2\omega_1^2
    - c_2\omega_2)}{\alpha (\kappa^2 - c_1^2)(c_2 - 2\omega_2)} \sqrt{E^2 - D^2 - \sum_i P_i^2} \ ,
\end{equation}
then the quantity,
\begin{equation}
    \tilde{E} - J_1 = \frac{2T_1k[2c_1\kappa(2\omega_1^2 - 2\omega_2^2 + c_2\omega_2)  -\omega_1(2c_1^2\omega_2
    + c_2\kappa^2 - 2\kappa^2\omega_2)]}{c_1c_2\kappa \omega_1}\sin\theta_1\cos\theta_1 \ ,
\end{equation}
is finite. This exactly adheres to the case of the dyonic giant magnon i.e. bound state of $J_2$ number of
giant magnons. In that case we also have $E$ and $J_1$ divergent, but $E- J_1$ finite and $J_2$ held fixed.
We can write the final dispersion relation in the form,
\begin{equation}
    \tilde{E} - J_1 = \sqrt{J_2^2 + f_2(\lambda)\sin^2\Big( \frac{\Delta\phi}{2}\Big)} \ ,
\end{equation}
where
again we have used the definition of effective 't Hooft coupling as before. Also $f_2(\lambda)$ remains
very complicated as in the previous case. But if we choose
$c_2\kappa = 2\kappa \omega_2 + c_1\omega_1$ then
we can reduce it to  $f_2(\lambda) = \frac{-3\lambda}{\pi^2}$ and the dispersion relation will become,
\begin{equation}
    \tilde{E} - J_1 = \sqrt{J_2^2 - \frac{3\lambda}{\pi^2}\sin^2\Big( \frac{\Delta\phi}{2}\Big)} \ .
\end{equation}
Once again one can see this relations exactly matches with the giant magnon dispersion 
relation obtained in \cite{Biswas:2011wu}.

\section{Rotating D1-strings on intersecting D5-D5' branes}
We shall now analyse the rotating D1-strings in the background of two stacks of
D5-branes intersecting on a line. More precisely, we have $N_1$ D5-branes extended along $(x_0, x_1, x_2, x_3, x_4, x_5)$
direction and $N_2$ D5-branes extended along $(x_0, x_1, x_6, x_7, x_8, x_9)$ directions, having exactly eight relatively 
transverse dimensions. These $N_1$ and $N_2$ have to be
large if we want the supergravity approximation to be valid, which constrain the regime of the coupling constant where the
solution is valid. For this reason, most of the literature works with the S dual configuration \cite{Itzhaki:2005tu}.
Nonetheless, it is a very interesting solution having the background of the form,
\begin{equation}
    ds^2 = (h_1h_2)^{-\frac{1}{2}} (-dt^2 + dx_1^2) + h_1^{-\frac{1}{2}} h_2^{\frac{1}{2}}\sum_{i=2}^5dx_i^2 +
    h_1^{\frac{1}{2}} h_2^{-\frac{1}{2}}\sum_{i=6}^9dx_i^2 \ ,
\end{equation}
together with a non-trivial dilaton,
\begin{equation}
    e^{2\phi} = h_1^{-1}h_2^{-1} \ ,
\end{equation}
We have to remember that there are two sets of transverse spaces for the intersecting brane configuration. For the 
set of branes having worldvolume lying on $(x_0, x_1, x_2, x_3, x_4, x_5)$, the transverse space is along 
$(x_6, x_7, x_8, x_9)$. We can measure the transverse radial direction in this case using $r_1^2 = \sum_{i=6}^9 x_i^2$.
Similarly for the second set of branes the transverse direction is parameterised by $r_2^2 = \sum_{i=2}^5 x_i^2$.
The harmonic functions for these set of branes are given by
\begin{equation}
h_1=1+\frac{k_1}{r_1^2}~~~ \text{and} ~~~h_2=1+\frac{k_2}{r_2^2}
\end{equation}
with $k_1=g_s N_1l_s^2$, $k_2=g_s N_2l_s^2$ and also
$\sum_{i=2}^5 dx_i^2= dr_2^2+r_2^2 d\Omega_2^2$, $\sum_{i=6}^9 dx_i^2= dr_1^2+r_1^2 d\Omega_1^2$.
For this configuration it is natural to talk about dual 3-form RR fluxes which like the previous section 
can actually be constructed from the volume form of the transverse spheres ($\Omega_1, \Omega_2$) of the
two sets of branes. Here we can parameterise the spheres as $d\Omega_1^2=d\theta_1^2 + \sin^2\theta_1d\phi_1^2 
+ \cos^2\theta_1 d\psi_1^2$ and
$d{\Omega}_2^2=d\theta_2^2 + \sin^2\theta_2d\phi_2^2 + \cos^2\theta_2 d\psi_2^2$. Calculating the RR fluxes, we get
\begin{eqnarray}
    \tilde{C}_{\phi_1 \psi_1}= 2k_1\sin^2\theta_1 \ , \>\>\>\> \tilde{C}_{\phi_2 \psi_2}= 2k_2\sin^2\theta_2 \ .
\end{eqnarray}
Now in the near-horizon $(r_1 \to 0, r_2 \to 0)$, we can probe the region near the brane intersections. In this limit
the harmonic functions approximate as $h_1 \approx \frac{k_1}{r_1^2},
    h_2 \approx \frac{k_2}{r_2^2}$. The metric will reduce as,
\begin{equation}
    ds^2 = \frac{r_1r_2}{\sqrt{k_1k_2}}\Big[-dt^2 + dx_1^2 + k_1\Big(\frac{dr_1^2}{r_1^2} +
    d\Omega_1^2\Big) +  k_2\Big(\frac{dr_2^2}{r_2^2} + d{\Omega}_2^2\Big)\Big] \ ,
\end{equation}

The dilaton in this limit becomes,
\begin{equation}
    e^{-\phi} = \sqrt{h_1h_2} = \frac{\sqrt{k_1k_2}}{r_1r_2} \ .
\end{equation}

Before proceeding further we consider both the stacks contains same (large) number of D5-branes i.e., $N_1=N_2=N$, which implies
$k_1=k_2=k$. With these simplification the metric,
dilaton and magnetic dual RR fields become,
\begin{eqnarray}
    ds^2 &=& \frac{r_1r_2}{k}\Big[-dt^2 + dx_1^2 + k\Big(\frac{dr_1^2}{r_1^2} + d\Omega_1^2\Big) +
    k\Big(\frac{dr_2^2}{r_2^2} + d{\Omega}_2^2\Big)\Big] \ , \nonumber \\
    e^{-\phi} &=& \frac{k}{r_1r_2} \ , \>\>\>\> \tilde{C}_{\phi_1,\psi_1}= 2k\sin^2\theta_1 \ ,
    \>\>\>\> \tilde{C}_{\phi_2,\psi_2}= 2k\sin^2\theta_2 \ .
\end{eqnarray}
 Further, on rescaling $t \to \sqrt{k}t$ and
$x_1 \to \sqrt{k}x_1$ and defining $\rho_1=\ln r_1$ and $\rho_2 = \ln r_2$, the metric and the dilaton becomes,
\begin{eqnarray}
     ds^2 &=& e^{\rho_1}e^{\rho_2}[-dt^2 + dx_1^2 + d\rho_1^2 +d\theta_1^2 + \sin^2\theta_1d\phi_1^2 +
     \cos^2\theta_1 d\psi_1^2 \nonumber \\ && + d\rho_2^2 +d\theta_2^2 + \sin^2\theta_2d\phi_2^2 + \cos^2\theta_2 d\psi_2^2]
     \ , \>\>\>\> e^{-\phi} = ke^{-\rho_1}e^{-\rho_2} \ ,
\end{eqnarray}
while the magnetic dual RR field remains unchanged. Now, the Lagrangian-density can be written as,
\begin{eqnarray}
    \mathcal{L} &=& -T_1e^{-\phi}\sqrt{-\det\>A_{\alpha \beta}} + \frac{T_1}{2} \epsilon^{\alpha\beta}
    \partial_{\alpha}X^{M}\partial_{\beta}X^{N} C_{MN}\nonumber \\ &=& -T_1k\Big[[-\partial_0t\partial_1t
    + \partial_0x_1\partial_1x_1 + \partial_0 \rho_1\partial_1 \rho_1 + \partial_0\theta_1\partial_1\theta_1
    + \sin^2\theta_1 \partial_0\phi_1\partial_1\phi_1 + \cos^2\theta_1 \partial_0\psi_1\partial_1\psi_1
    \nonumber \\ && + \partial_0 \rho_2\partial_1 \rho_2 + \partial_0\theta_2\partial_1\theta_2 +
    \sin^2\theta_2 \partial_0\phi_2\partial_1\phi_2 + \cos^2\theta_2 \partial_0\psi_2\partial_1\psi_2]^2
    -[-(\partial_0t)^2 + (\partial_0x_1)^2 \nonumber \\ && + (\partial_0 \rho_1)^2 + (\partial_0\theta_1)^2
    + \sin^2\theta_1 (\partial_0\phi_1)^2 + \cos^2\theta_1 (\partial_0\psi_1)^2 + (\partial_0 \rho_2)^2
    + (\partial_0\theta_2)^2 \nonumber \\ && + \sin^2\theta_2 (\partial_0\phi_2)^2 + \cos^2\theta_2
    (\partial_0\psi_2)^2][-(\partial_1t)^2 + (\partial_1x_2)^2 + (\partial_1 \rho_1)^2 + (\partial_1\theta_1)^2
    + \sin^2\theta_1 (\partial_1\phi_1)^2 \nonumber \\ && + \cos^2\theta_1 (\partial_1\psi_1)^2 +
    (\partial_1 \rho_2)^2 + (\partial_1\theta_2)^2 + \sin^2\theta_2 (\partial_1\phi_2)^2 + \cos^2\theta_2
    (\partial_1\psi_2)^2]\Big]^{\frac{1}{2}} \nonumber \\ && + 2T_1k\sin^2\theta_1(\partial_0\phi_1\partial_1\psi_1
    - \partial_0\psi_1\partial_1\phi_1) + 2T_1k\sin^2\theta_2(\partial_0\phi_2\partial_1\psi_2 -
    \partial_0\psi_2\partial_1\phi_2) \ .
\end{eqnarray}
 Before solving the Euler-Lagrange equations, we assume the following rotating string ansatz,
\begin{eqnarray}
   && t= \kappa \xi^0 \ , \>\>\> x_1 = v\xi^0 \ , \>\>\> \rho_i = m_i\xi^0 \ , \>\>\>
   \theta_i = \theta_i(\xi^1) \ , \nonumber \\ && \phi_i = \nu_i\xi^0 + \xi^1 \ ,
   \>\>\> \psi_i = \omega_i\xi^0 + \psi_i(\xi^1) \ , \>\>\> i=1,2 \ .
\end{eqnarray}
As we did in the last section also, these are basically generalizations of rotating F-string solutions
which can be shown to be consistent with the equations of motion. 
Now by solving the equation of motion for $t$, we get,
\begin{equation}
    \frac{\kappa(\nu_1\sin^2\theta_1 + \omega_1 \partial_1\psi_1\cos^2\theta_1 + \nu_2\sin^2\theta_2 +
    \omega_2 \partial_1\psi_2\cos^2\theta_2 )}{\sqrt{B}} = c_1 \ , \label{teqn}
\end{equation}
where for convenience we have defined
\begin{eqnarray}
    B &=& [\nu_1\sin^2\theta_1 + \omega_1 \partial_1\psi_1\cos^2\theta_1 + \nu_2\sin^2\theta_2 +
    \omega_2 \partial_1\psi_2\cos^2\theta_2]^2 - [-\alpha^2 + \nu_1^2\sin^2\theta_1 + \nonumber \\
    && \omega_1^2\cos^2\theta_1 + \nu_2^2\sin^2\theta_2 + \omega_2^2\cos^2\theta_2][(\partial_1\theta_1)^2
    + (\partial_1\theta_2)^2 + \sin^2\theta_1 \nonumber \\ && + \sin^2\theta_2 + \cos^2\theta_1(\partial_1\psi_1)^2
    + \cos^2\theta_2(\partial_1\psi_2)^2]
\end{eqnarray}
and $\alpha^2 = \kappa^2 - v^2 - m_1^2 - m_2^2$ and $c_1$ are just constants.

Now solving the equations of motion for $\phi_1$ and $\psi_1$, we get,
\begin{eqnarray}
 \frac{\sin^2\theta_1}{\sqrt{B}} && \Big[\omega_1(\omega_1-\nu_1\partial_1\psi_1)\cos^2\theta_1 +
 \nu_2(\nu_2-\nu_1)\sin^2\theta_2 \nonumber \\ && + \omega_2(\omega_2 - \nu_1\partial_1\psi_2)\cos^2\theta_2
 - \alpha^2\Big] - 2\omega_1\sin^2\theta_1 = c_2 \ , \label{phi1} \\ \frac{\cos^2\theta_1}{\sqrt{B}} &&
 \Big[\nu_1(\nu_1\partial_1\psi_1-\omega_1)\sin^2\theta_1 + \nu_2(\nu_2\partial_1\psi_1-\omega_1)\sin^2\theta_2
 \nonumber \\ && + \omega_2(\omega_2\partial_1\psi_1 - \omega_1\partial_1\psi_2)\cos^2\theta_2 -
 \alpha^2\partial_1\psi_1\Big] + 2\nu_1\sin^2\theta_1 = c_3 \ . \label{psi1}
\end{eqnarray}
Again, solving for $\phi_2$ and $\psi_2$, we get,
\begin{eqnarray}
 \frac{\sin^2\theta_2}{\sqrt{B}} && \Big[\nu_1(\nu_1-\nu_2)\sin^2\theta_1 + \omega_1(\omega_1
 -\nu_2\partial_1\psi_1)\cos^2\theta_1 \nonumber \\ && + \omega_2(\omega_2 - \nu_2\partial_1
 \psi_2)\cos^2\theta_2 - \alpha^2\Big] - 2\omega_2\sin^2\theta_2 = c_4 \ , \label{phi2} \\
 \frac{\cos^2\theta_2}{\sqrt{B}} && \Big[\nu_1(\nu_1\partial_1\psi_2-\omega_2)\sin^2\theta_1
 + \omega_1(\omega_1\partial_1\psi_2-\omega_2\partial_1\psi_1)\cos^2\theta_1 \nonumber \\
 && + \nu_2(\nu_2\partial_1\psi_2 - \omega_2)\sin^2\theta_2 - \alpha^2\partial_1\psi_2\Big]
 + 2\nu_2\sin^2\theta_2 = c_5 \ . \label{psi2}
\end{eqnarray}
Solving (\ref{psi1}) we get,
\begin{eqnarray}
   && [c_1\nu_1^2\sin^2\theta_1 + c_1\nu_2^2\sin^2\theta_2 + c_1\omega_2^2\cos^2\theta_2 + 2\kappa \nu_1\omega_1\sin^2\theta_1 
   - c_1\alpha^2 - c_3\kappa\omega_1]\partial_1\psi_1\cos^2\theta_1 \nonumber \\ && = (c_1\omega_1\cos^2\theta_1 -
   2\kappa\nu_1\sin^2\theta_1 + c_3\kappa)(\omega_2\partial_1\psi_2\cos^2\theta_2 + \nu_1\sin^2\theta_1 + \nu_2\sin^2\theta_2)
   \ . \label{eqn1}
\end{eqnarray}
Again, solving (\ref{psi2}) we get,
\begin{eqnarray}
    && [c_1\nu_1^2\sin^2\theta_1 + c_1\nu_2^2\sin^2\theta_2 + c_1\omega_1^2\cos^2\theta_1 + 2\kappa \nu_2\omega_2\sin^2\theta_2 
    - c_1\alpha^2 - c_5\kappa\omega_2]\partial_1\psi_2\cos^2\theta_2 \nonumber \\ && = (c_1\omega_2\cos^2\theta_2 - 
    2\kappa\nu_2\sin^2\theta_2 + c_5\kappa)(\omega_1\partial_1\psi_1\cos^2\theta_1 + \nu_1\sin^2\theta_1 + \nu_2\sin^2\theta_2)
    \ . \label{eqn2}
\end{eqnarray}
And finally from equation (\ref{teqn}), we have,
\begin{eqnarray}
    (\partial_1\theta_1)^2 + (\partial_1\theta_1)^2 && = \frac{(c_1^2 - \kappa^2)(\nu_1\sin^2\theta_1 +
    \omega_1\partial_1\psi_1\cos^2\theta_1 + \nu_2\sin^2\theta_2 + \omega_2\partial_1\psi_2\cos^2\theta_2)^2}
    {c_1^2(-\alpha^2 + \nu_1^2\sin^2\theta_1 + \omega_1^2\cos^2\theta_1 + \nu_2^2\sin^2\theta_2 +
    \omega_2^2\cos^2\theta_2)} \nonumber \\ && -\sin^2\theta_1 - \sin^2\theta_2 - \cos^2\theta_1
    (\partial_1\psi_1)^2 - \cos^2\theta_2 (\partial_1\psi_2)^2 \ . \label{theta1}
\end{eqnarray}

Equation (\ref{theta1}) have two variables $\partial_1\theta_1$ and $\partial_1\theta_2$, it is quite difficult
to find the solutions of both $\partial_1\theta_1$ and $\partial_1\theta_2$ simultaneously from the form of 
the equation (\ref{theta1}). For simplicity we will now consider $\theta_1=\theta_2=\theta$, which will confine 
the string on both the spheres but with restriction that both the sphere will have the same value of $\theta$.
This will make our equations more tractable.

Also, we define the following quantities for our convenience,
\begin{eqnarray}
    a = c_1(\nu_1^2 + \nu_2^2)\sin^2\theta - c_1\alpha^2 \ ,  \>\>\> && b = c_1\omega_2^2\cos^2\theta
    + 2\kappa \nu_1\omega_1 \sin^2\theta - c_3\kappa\omega_1 \ , \nonumber \\ c = c_1\omega_1^2\cos^2\theta
    + 2\kappa \nu_2\omega_2 \sin^2\theta - c_5\kappa\omega_2 \ , \> && d= (\nu_1+\nu_2)\sin^2\theta \ ,
    \nonumber \\ e = c_1\omega_1\cos^2\theta - 2\kappa\nu_1\sin^2\theta + c_3\kappa \ , \>\>\> && f =
    c_1\omega_2\cos^2\theta - 2\kappa\nu_2\sin^2\theta + c_5\kappa \ ,
\end{eqnarray}
then equation (\ref{eqn1}) and (\ref{eqn2}) can be written as,
\begin{eqnarray}
    (a+b)\partial_1\psi_1\cos^2\theta &=& e(\omega_2\partial_1\psi_2\cos^2\theta +d) \ , \nonumber \\
    (a+c)\partial_1\psi_2\cos^2\theta &=& f(\omega_1\partial_1\psi_1\cos^2\theta +d) \ .
\end{eqnarray}
By solving the above equations we get,
\begin{eqnarray}
    \partial_1\psi_1 &=& \frac{ed(a+c+\omega_2f)}{\cos^2\theta[(a+b)(a+c)-\omega_1\omega_2ef]} \ ,
    \nonumber \\ \partial_1\psi_2 &=& \frac{fd(a+b+\omega_1e)}{\cos^2\theta[(a+b)(a+c)-\omega_1\omega_2ef]} \ .
\end{eqnarray}
Again using these results equation (\ref{theta1}) reduces to,
\begin{eqnarray}
    (\partial_1\theta)^2 &=& \frac{(\kappa^2 - c_1^2)}{2c_1^2} \frac{[(\nu_1 + \nu_2)\sin^2\theta +
    (\omega_1\partial_1\psi_1 + \omega_2\partial_2\psi_2)\cos^2\theta]^2}{[\alpha^2 - (\nu_1^2 + \nu_2^2)
    \sin^2\theta - (\omega_1^2 + \omega_2^2)\cos^2\theta]} \nonumber \\ && - \sin^2\theta -\frac{1}{2}
    [(\partial_1\psi_1)^2 + (\partial_1\psi_2)^2]\cos^2\theta \ . \label{theta2}
\end{eqnarray}
The above set of equations of motion appear to be generalizations of the ones discussed in the last section for single
stack of D5 branes. We will exactly follow the way we found out the single spike and giant magnon solutions for the
last section. Namely we will impose the similar kind of conditions on $\partial_1\theta$, $\partial_1\psi_1$ and
$\partial_1\psi_2$ and define the conserved charges in each case separately. 
\subsection{Single Spike-like solution}
For a generalised spike solution, we impose the condition on the derivatives
\begin{equation}
\partial_1\theta \to 0~~~ \text{as} ~~~\theta \to \frac{\pi}{2},
\end{equation}
Implementing the conditions, from equation (\ref{theta2}), we get,
\begin{eqnarray}
    c_1 = \frac{\kappa(\nu_1 + \nu_2)}{\alpha_1} \ , \>\>\> \text{where} \>\>\>
    \alpha_1 = \sqrt{2\alpha^2 - (\nu_1 - \nu_2)^2} \ .
\end{eqnarray}
Again, follwing our earlier discussion we will demand that the derivatives  $\partial_1\psi_1$ and $\partial_2\psi_2$
have to be regular in the limit $\theta \to \frac{\pi}{2}$.
This leads us to the relations between the constants $c_3=2\nu_1$ and $c_5=2\nu_2$ respectively.
Using these relations, we get,
\begin{eqnarray}
    \partial_1\psi_1 &=& \frac{(\nu_1 + \nu_2)(c_1\omega_1 + 2\kappa\nu_1)\sin^2\theta}
    {c_1(\nu_1^2 + \nu_2^2)\sin^2\theta - 2\kappa\alpha_2\cos^2\theta - c_1\alpha^2} \ ,
    \nonumber \\ \partial_1\psi_2 &=& \frac{(\nu_1 + \nu_2)(c_1\omega_2 + 2\kappa\nu_2)
    \sin^2\theta}{c_1(\nu_1^2 + \nu_2^2)\sin^2\theta - 2\kappa\alpha_2\cos^2\theta - c_1\alpha^2} \ ,
\end{eqnarray}
where $\alpha_2 = \nu_1\omega_1 + \nu_2\omega_2$.
Similarly, equation (\ref{theta2}) reduces to,
\begin{equation}
    \partial_1\theta = \sqrt{\frac{a_1}{2}}\frac{\kappa\sin\theta\cos\theta \sqrt{\sin^2\theta -
    \sin^2\theta_0}}{\alpha_1[c_1(\nu_1^2 + \nu_2^2)\sin^2\theta - 2\kappa \alpha_2\cos^2\theta -
    c_1\alpha^2]} \ ,
\end{equation}
where
\begin{equation}
\sin\theta_0 = \frac{2\alpha_1\alpha_2 + (\nu_1 + \nu_2)\alpha^2}{\sqrt{\frac{a_1}{2}}}
\end{equation}
and
\begin{eqnarray}
a_1 &=& 2\alpha^2(\nu_1^2 + \nu_2^2)(\nu_1 + \nu_2)^2 + 8\alpha_1^2\alpha_2^2 + 8\alpha_1\alpha_2(\nu_1
+ \nu_2)(\nu_1^2 + \nu_2^2) - 4\alpha_1\alpha_2(\nu_1 + \nu_2)^3 \nonumber\\
&-& \alpha_1^2(\nu_1 + \nu_2)(\omega_1^2 +
\omega_2^2 + 4\nu_1^2 + \nu_2^2).
\end{eqnarray}
Now, like the previous section we can solve for the string profile by integrating
the $\theta$ equation. We will not present the cumbersome expressions here for brevity.

Let us now define the conserved charges corresponding to the various isometries as given by,
\begin{eqnarray}
    E &=& -2T_1\int\frac{\partial\mathcal{L}}{\partial(\partial_0t)} \frac{d\theta}{\partial_1\theta}
    = \frac{4T_1k\kappa(\nu_1+\nu_2)(\nu_1^2 + \nu_2^2 - \alpha^2)}{\alpha_1 \sqrt{\frac{a_1}{2}}}
    \int_{\theta_0}^{\frac{\pi}{2}} \frac{\sin\theta d\theta}{\cos\theta \sqrt{\sin^2\theta -
    \sin^2\theta_0}} \ , \nonumber \\ P &=& 2T_1\int\frac{\partial\mathcal{L}}{\partial(\partial_0x_1)}
    \frac{d\theta}{\partial_1\theta} = \frac{4T_1k v(\nu_1+\nu_2)(\nu_1^2 + \nu_2^2 - \alpha^2)}{\alpha_1
    \sqrt{\frac{a_1}{2}}} \int_{\theta_0}^{\frac{\pi}{2}} \frac{\sin\theta d\theta}{\cos\theta
    \sqrt{\sin^2\theta - \sin^2\theta_0}} \ , \nonumber \\ D_1 &=& 2T_1\int\frac{\partial\mathcal{L}}
    {\partial(\partial_0\rho_1)} \frac{d\theta}{\partial_1\theta} = \frac{4T_1k m_1(\nu_1+\nu_2)
    (\nu_1^2 + \nu_2^2 - \alpha^2)}{\alpha_1 \sqrt{\frac{a_1}{2}}} \int_{\theta_0}^{\frac{\pi}{2}}
    \frac{\sin\theta d\theta}{\cos\theta \sqrt{\sin^2\theta - \sin^2\theta_0}} \ , \nonumber \\
    D_2 &=& 2T_1\int\frac{\partial\mathcal{L}}{\partial(\partial_0\rho_2)} \frac{d\theta}{\partial_1\theta}
    = \frac{4T_1k m_2(\nu_1+\nu_2)(\nu_1^2 + \nu_2^2 - \alpha^2)}{\alpha_1 \sqrt{\frac{a_1}{2}}}
    \int_{\theta_0}^{\frac{\pi}{2}} \frac{\sin\theta d\theta}{\cos\theta \sqrt{\sin^2\theta -
    \sin^2\theta_0}} \ . \nonumber \\
\end{eqnarray}
Like the previous section, the charges $D_1$ and $D_2$ are associated with shifts in $\ln r_1$ and $\ln r_2$,
which again are symmetries of the D1 string action.
All these above quantities can be shown to diverge. But, using these expressions we can define a new divergent quantity,
\begin{equation}
    \sqrt{E^2 - P^2 - D_1^2 - D_2^2} = \frac{4T_1k \alpha(\nu_1+\nu_2)(\nu_1^2 + \nu_2^2 - \alpha^2)}
    {\alpha_1 \sqrt{\frac{a_1}{2}}} \int_{\theta_0}^{\frac{\pi}{2}} \frac{\sin\theta d\theta}
    {\cos\theta \sqrt{\sin^2\theta - \sin^2\theta_0}} \ .
\end{equation}
The angle deficit $\Delta\phi = 2\int \frac{d\theta}{\partial_1\theta}$ is given by,
\begin{equation}
    \Delta\phi = \frac{2\alpha_1}{\kappa\sqrt{\frac{h_1}{2}}} \Big[ c_1(\nu_1^2 + \nu_2^2 - \alpha^2)
    \int_{\theta_0}^{\frac{\pi}{2}} \frac{\sin\theta d\theta}{\cos\theta \sqrt{\sin^2\theta - \sin^2\theta_0}}
    - (c_1\alpha^2 + 2\kappa \alpha_2) \int_{\theta_0}^{\frac{\pi}{2}} \frac{\sin\theta \cos\theta d\theta}
    {\sqrt{\sin^2\theta - \sin^2\theta_0}}\Big] \ .
\end{equation}
This is also divergent, but we can regularize it by removing the divergent part,
\begin{equation}
    (\Delta\phi)_{reg} = \Delta\phi - \frac{\alpha_1}{2T_1k\alpha}\sqrt{E^2-P^2 - D_1^2 - D_2^2} =
    -\cos^{-1}(\sin\theta_0) \ ,
\end{equation}
which implies $\sin\theta_0 = \cos (\frac{(\Delta\phi)_{reg}}{2})$. Angular momenta $J_1 =
2T_1\int\frac{\partial\mathcal{L}}{\partial(\partial_1\phi_1)} \frac{d\theta}{\partial_1\theta}$
and $J_2 = 2T_1\int\frac{\partial\mathcal{L}}{\partial(\partial_1\phi_2)} \frac{d\theta}{\partial_1\theta}$
also diverge. Regularised values of $J_1$ and $J_2$ are given by,
\begin{eqnarray}
    (J_1)_{reg} &=& J_1 - \frac{2\alpha_1(\nu_1 + \nu_2)\omega_1 + 4\alpha_1^2\nu_1 + (\nu_1 - \nu_2)
    (\nu_1^2 + \nu_2^2 - \alpha^2)}{2\alpha(\nu_1^2 + \nu_2^2 - \alpha^2)} \sqrt{E^2 - P^2 - D_1^2 - D_2^2}
    \nonumber \\ && = \frac{2T_1k(\nu_1 + \nu_2)}{\alpha_1\sqrt{\frac{a_1}{2}}}[2\alpha_1\nu_2(\omega_2 - \omega_1)
    + 2\nu_1(\alpha^2 - 2\alpha_1^2) - (\nu_1 - \nu_2)(\nu_1^2 + \nu_2^2)]\cos\theta_0 \ , \nonumber \\
    (J_2)_{reg} &=& J_2 - \frac{2\alpha_1(\nu_1 + \nu_2)\omega_2 + 4\alpha_1^2\nu_1 + (\nu_2 - \nu_1)
    (\nu_1^2 + \nu_2^2 - \alpha^2)}{2\alpha(\nu_1^2 + \nu_2^2 - \alpha^2)} \sqrt{E^2 - P^2 - D_1^2 - D_2^2}
    \nonumber \\ && = \frac{2T_1k(\nu_1 + \nu_2)}{\alpha_1\sqrt{\frac{a_1}{2}}}[2\alpha_1\nu_1(\omega_1 - \omega_2)
    + 2\nu_2(\alpha^2 - 2\alpha_1^2) + (\nu_1 - \nu_2)(\nu_1^2 + \nu_2^2)]\cos\theta_0 \ . \nonumber \\
\end{eqnarray}

Again the angular momenta $K_1 = 2T_1\int\frac{\partial\mathcal{L}}{\partial(\partial_1\psi_1)}
\frac{d\theta}{\partial_1\theta}$ and $K_2 = 2T_1\int\frac{\partial\mathcal{L}}{\partial(\partial_1\psi_2)}
\frac{d\theta}{\partial_1\theta}$ also diverges. Regularised $K_1$ and $K_2$ are given by,
\begin{eqnarray}
    (K_1)_{reg} &=& K_1 + \frac{\alpha_1}{\alpha} \sqrt{E^2 - P^2 - D_1^2 - D_2^2} \nonumber \\ && =
    \frac{2T_1k}{\sqrt{\frac{a_1}{2}}}[4\alpha_1\alpha_2 - (\nu_1 + \nu_2)\omega_1\alpha_1 - 2\nu_2^2
    (\nu_1^2 - \nu_2^2)]\cos\theta_0 \ , \nonumber \\ (K_2)_{reg} &=& K_2 - \frac{\alpha_1}{\alpha}
    \sqrt{E^2 - P^2 - D_1^2 - D_2^2} \nonumber \\ && = \frac{2T_1k}{\sqrt{\frac{a_1}{2}}}[4\alpha_1\alpha_2 -
    (\nu_1 + \nu_2)\omega_2\alpha_1 - 2\nu_1^2(\nu_2^2 - \nu_1^2)]\cos\theta_0 \ ,
\end{eqnarray}
Now, defining $J_{reg} = (J_1)_{reg} + (J_2)_{reg}$ and $K_{reg} = (K_1)_{reg} - (K_2)_{reg}$,
we find that they satisfy a generalized dispersion relation of form,
\begin{equation}
    J_{reg} = \sqrt{K_{reg}^2 + f_3(\lambda)\sin^2 \Big(\frac{(\Delta\phi)_{reg}}{2}\Big)} \ ,
\end{equation}
where $f_3(\lambda) = \frac{2\lambda}{\pi^2}\frac{(\nu_1 + \nu_2)^2}{a_1\alpha_1^2}[\{2\alpha_1(\omega_2 - \omega_1)
(\nu_2 - \nu_1) + 2(\nu_1 + \nu_2)(\alpha^2 - 2\alpha_1^2)\}^2 - \alpha_1^2 \{(\omega_2 - \omega_1)\alpha_1 +
(\nu_2^2 - \nu_1^2)\}^2]$. 

\subsection{Giant Magnon-like solution}
Here we use the opposite condition on the equations of motion, i.e. we impose that $\partial_1\theta$,
$\partial_1\psi_1$ and $\partial_1\psi_2$ diverges as $\theta \to \frac{\pi}{2}$. Imposing the first one, we get 
the following condition,
\begin{equation}
    \alpha^2 = \nu_1^2 + \nu_2^2 ~~~
\end{equation}
we can not determine the values of $c_3$ and $c_5$ by this condition alone. As we
have found in the previous cases, we put $c_3=2\nu_1$ and $c_5=2\nu_2$ by hand. Using these
values of constants, we get,
\begin{eqnarray}
    \partial_1\psi_1 &=& -\frac{(\nu_1 + \nu_2)(c_1\omega_1 + 2\kappa\nu_1)\sin^2\theta}{\beta \cos^2\theta}
    \ , \nonumber \\ \partial_1\psi_2 &=& -\frac{(\nu_1 + \nu_2)(c_1\omega_2 + 2\kappa\nu_2)\sin^2\theta}
    {\beta \cos^2\theta} \ ,
\end{eqnarray}
where $\beta = c_1(\nu_1^2 + \nu_2^2) + 2\kappa \alpha_2$. Also, equation (\ref{theta2}) reduces to,
\begin{equation}
    \partial_1\theta = \frac{\sin\theta \sqrt{\sin^2\theta - \sin^2\theta_1}}{\sin\theta_1\cos\theta} \ ,
\end{equation}
where $\sin\theta_1 = \frac{\beta}{\sqrt{\frac{a_2}{2}}}$ and $a_2 = 2\beta^2 - (\nu_1 + \nu_2)^2
[(3\kappa^2 + c_1^2)(\nu_1^2 + \nu_2^2) + \kappa^2(\omega_1^2 + \omega_2^2 + 4c_2\kappa \alpha_2)]$.

In this case, the conserved charges are given by,
\begin{eqnarray}
    E &=&  \frac{2T_1k(\nu_1+\nu_2)(c_1^2 - \kappa^2)}{\sqrt{\frac{a_2}{2}}} \int^{\theta_1}_{\frac{\pi}{2}}
    \frac{\sin\theta d\theta}{\cos\theta \sqrt{\sin^2\theta - \sin^2\theta_1}} \ , \nonumber \\ P &=&
    \frac{2T_1k v(\nu_1+\nu_2)(c_1^2  - \kappa^2)}{\kappa \sqrt{\frac{a_2}{2}}} \int^{\theta_1}_{\frac{\pi}{2}}
    \frac{\sin\theta d\theta}{\cos\theta \sqrt{\sin^2\theta - \sin^2\theta_1}} \ , \nonumber \\ D_1 &=&
    \frac{2T_1k m_1(\nu_1+\nu_2)(c_1^2 - \kappa^2)}{\kappa \sqrt{\frac{a_2}{2}}} \int^{\theta_1}_{\frac{\pi}{2}}
    \frac{\sin\theta d\theta}{\cos\theta \sqrt{\sin^2\theta - \sin^2\theta_1}} \ , \nonumber \\ D_2 &=&
    \frac{2T_1k m_2(\nu_1+\nu_2)(c_1^2 - \kappa^2)}{\kappa \sqrt{\frac{a_2}{2}}} \int^{\theta_1}_{\frac{\pi}{2}}
    \frac{\sin\theta d\theta}{\cos\theta \sqrt{\sin^2\theta - \sin^2\theta_1}} \ .
\end{eqnarray}
All these quantities diverge and  we again define the quantity,
\begin{equation}
    \sqrt{E^2 - P^2 - D_1^2 - D_2^2} = \frac{2T_1k \alpha(\nu_1+\nu_2)(c_1^2 -  \kappa^2)}{\kappa \sqrt{\frac{a_2}{2}}}
    \int^{\theta_1}_{\frac{\pi}{2}} \frac{\sin\theta d\theta}{\cos\theta \sqrt{\sin^2\theta - \sin^2\theta_1}} \ .
\end{equation}
The angle deficit is finite and is given by,
\begin{equation}
    \Delta\phi = 2\sin\theta_1\int_{\frac{\pi}{2}}^{\theta_1} \frac{\cos\theta d\theta}{\sin\theta\sqrt{
\sin^2\theta - \sin^2\theta_1}} = 2\cos^{-1}(\sin\theta_1) \ ,
\end{equation}
which implies $\sin\theta_1 = \cos (\frac{\Delta\phi}{2})$. Again, the angular momenta $J_1$ and $J_2$  are given by,
\begin{eqnarray}
    J_1 &=& \frac{2T_1k}{\kappa \sqrt{\frac{a_2}{2}}} \Big[ [c_1\beta + (\nu_1 + \nu_2)\{2\kappa(c_1\omega_1 +
    2\kappa\nu_1) - \nu_1(c_1^2 - \kappa^2)\}]\int_{\frac{\pi}{2}}^{\theta_1} \frac{\sin\theta\cos\theta d\theta}
    {\sqrt{\sin^2\theta - \sin^2\theta_1}} \nonumber \\ && + (\nu_1 + \nu_2)[\nu_1(c_1^2 - \kappa^2) -
    2\kappa(c_1\omega_1+ 2\kappa\nu_1)] \int_{\frac{\pi}{2}}^{\theta_1} \frac{\sin\theta d\theta}
    {\cos\theta\sqrt{\sin^2\theta - \sin^2\theta_1}} \Big] \ ,  \\ J_2 &=& \frac{2T_1k}{\kappa \sqrt{\frac{a_2}{2}}}
    \Big[ [c_1\beta + (\nu_1 + \nu_2)\{2\kappa(c_1\omega_2 + 2\kappa\nu_2) - \nu_2(c_1^2 - \kappa^2)\}]
    \int_{\frac{\pi}{2}}^{\theta_1} \frac{\sin\theta\cos\theta d\theta}{\sqrt{\sin^2\theta - \sin^2\theta_1}}
    \nonumber \\ && + (\nu_1 + \nu_2)[\nu_2(c_1^2 - \kappa^2) - 2\kappa(c_1\omega_2 + 2\kappa\nu_2)]
    \int_{\frac{\pi}{2}}^{\theta_1} \frac{\sin\theta d\theta}{\cos\theta\sqrt{\sin^2\theta - \sin^2\theta_1}} \Big] \ ,
\end{eqnarray}

But, the angular momenta $K_1$ and $K_2$ are finite and are given by,
\begin{eqnarray}
    K_1 &=&  \frac{2T_1k}{\sqrt{\frac{a_2}{2}}}[(\nu_1 + \nu_2)(\omega_1\kappa + 2c_1\nu_1) + 2\beta]\cos\theta_1 \
    , \nonumber \\ K_2 &=&  \frac{2T_1k}{\sqrt{\frac{a_2}{2}}}[(\nu_1 + \nu_2)(\omega_2\kappa + 2c_1\nu_2) + 2\beta]
    \cos\theta_1 \ ,
\end{eqnarray}
Now, defining $J = J_1 + J_2$, $K = K_1 - K_2$, and
\begin{equation}
    \tilde{E} = \frac{(\nu_1 - \nu_2)(c_1^2 - 5\kappa^2) + 2c_1\kappa(\omega_1 - \omega_2)}{\alpha(c_1^2 - \kappa^2)}
    \sqrt{E^2 - P^2 - D_1^2 - D_2^2} \ ,
\end{equation}
we find that they satisfy a dispersion relation of form,
\begin{equation}
   \tilde{E} - J = \sqrt{K^2 + f_4(\lambda)\sin^2 \Big(\frac{\Delta\phi}{2}\Big)} \ ,
\end{equation}
where $f_4(\lambda) = \frac{2\lambda}{\pi^2}\frac{(\nu_1 + \nu_2)^2}{a_2\kappa^2}[\{(\nu_2 - \nu_1)(c_1^2 - 5\kappa^2)
+ 2c_1\kappa(\omega_1 - \omega_2)\}^2 - \kappa^2 \{\kappa(\omega_1 - \omega_2) + 2c_1(\nu_1 + \nu_2)\}^2]$.
Again, it is noteworthy that the same form of generalised giant magnon solution was found for a rotating F-string 
in \cite{Biswas:2012wu}.

\section{Conclusions}
In this paper, we have studied various solutions of the D-string
equations of motion in various curved backgrounds. First we have
studied string equations of motion, of a bound state of
oscillating D1 strings and F-strings with non-trivial gauge field
on the D1 worldvolume, in the recently found $AdS_3$ background
with mixed fluxes. One of the interesting outcome of this solution
is that the periodically expanding and contracting $(1, n)$ string
has a possibility of reaching the boundary of $AdS_3$ in finite time
in contrast to the probe D-string motion in the WZW model with
only NS-NS fluxes \cite{Bachas:2000fr}. Further we have studied the
DBI equations of motion of the D1-string in D5-brane background
and found out two classes of solutions corresponding to the giant
magnon and single spike solutions of the D-string. Also, we have
studied the rigidly rotating D1-string solution in the
intersecting D5-brane background and compared our results with the
existing S-dual configurations. These solutions are basically 
generalisations of the usual single spike and giant magnon solutions
for a D1-string. However, the physical interpretation
of such string states remain elusive from us as of now. 

\section*{Acknowledgements}
 It is a pleasure to thank Kamal L. Panigrahi for discussions and important
 suggestions on the manuscript. The authors are indebted to S. Pratik Khastgir
 for various fruitful discussions. AB would like to thank Pulastya
Parekh and Abhishake Sadhukhan for their comments. He would also like to thank
the organisers of National String Meeting 2015 for kind hospitality at IISER 
Mohali, where a part of this work was done.


\begin{thebibliography}{}
 \bibitem{Maldacena:1997re}
  J.~M.~Maldacena,
  ``The Large N limit of superconformal field theories and supergravity,''
  Int.\ J.\ Theor.\ Phys.\  {\bf 38}, 1113 (1999)
  [Adv.\ Theor.\ Math.\ Phys.\  {\bf 2}, 231 (1998)]
  [hep-th/9711200].

\bibitem{Witten:1998qj}
  E.~Witten,
  ``Anti-de Sitter space and holography,''
  Adv.\ Theor.\ Math.\ Phys.\  {\bf 2}, 253 (1998)
  [hep-th/9802150].


\bibitem{Gubser:1998bc}
  S.~S.~Gubser, I.~R.~Klebanov and A.~M.~Polyakov,
  ``Gauge theory correlators from noncritical string theory,''
  Phys.\ Lett.\ B {\bf 428}, 105 (1998)
  [hep-th/9802109].



\bibitem{Tseytlin:2004xa}
  A.~A.~Tseytlin,
  ``Semiclassical strings and AdS/CFT,''
  hep-th/0409296.



\bibitem{Balog:1988jb}
  J.~Balog, L.~O'Raifeartaigh, P.~Forgacs and A.~Wipf,
  ``Consistency of String Propagation on Curved Space-Times: An SU(1,1) Based Counterexample,''
  Nucl.\ Phys.\ B {\bf 325}, 225 (1989).


\bibitem{Banados:1992wn}
  M.~Banados, C.~Teitelboim and J.~Zanelli,
  ``The Black hole in three-dimensional space-time,''
  Phys.\ Rev.\ Lett.\  {\bf 69}, 1849 (1992)
  [hep-th/9204099].

\bibitem{Banados:1992gq}
  M.~Banados, M.~Henneaux, C.~Teitelboim and J.~Zanelli,
  ``Geometry of the (2+1) black hole,''
  Phys.\ Rev.\ D {\bf 48}, 1506 (1993)
  [Phys.\ Rev.\ D {\bf 88}, 069902 (2013)]
  [gr-qc/9302012].

\bibitem{klim}
C.~Klim{\v c}{\'\i}k and P.~{\v S}evera,
``Open strings and D-branes in WZNW models",
Nucl.\ Phys.\  {\bf B488}, 653 (1997)
[hep-th/9609112].

\bibitem{as}
A.Y.~Alekseev and V.~Schomerus,
``D-branes in the WZW model",
Phys.\ Rev.\  {\bf D60}, 061901 (1999)
[hep-th/9812193].

\bibitem{fffs}
G.~Felder, J.~Frohlich, J.~Fuchs and C.~Schweigert,
``The geometry of WZW branes",
J.\ Geom.\ Phys.\  {\bf 34}, 162 (2000)
[hep-th/9909030].

\bibitem{sta}
S.~Stanciu,
``D-branes in group manifolds",
JHEP {\bf 0001}, 025 (2000)
[hep-th/9909163].
%

\bibitem{ars}
A.Y.~Alekseev, A.~Recknagel and V.~Schomerus,
``Non-commutative world-volume geometries:
branes on $SU(2)$ and fuzzy  spheres",
JHEP {\bf 9909}, 023 (1999)
[hep-th/9908040].

\bibitem{bds}
C.~Bachas, M.R.~Douglas and C.~Schweigert,
``Flux stabilization of D-branes",
JHEP {\bf 0005}, 048 (2000)
[hep-th/0003037].

\bibitem{paw}
J.~Pawelczyk,
``$SU(2)$ WZW D-branes and their non-commutative geometry from DBI
action",
JHEP {\bf 0008}, 006 (2000)
[hep-th/0003057].


\bibitem{ars2}
A.Y.~Alekseev, A.~Recknagel and V.~Schomerus,
``Brane dynamics in background fluxes and non-commutative geometry",
JHEP {\bf 0005}, 010 (2000)
[hep-th/0003187].

\bibitem{Bachas:2000fr}
  C.~Bachas and M.~Petropoulos,
  ``Anti-de Sitter D-branes,''
  JHEP {\bf 0102}, 025 (2001)
  [hep-th/0012234].


\bibitem{Kluson:2007fr}
  J.~Kluson and K.~L.~Panigrahi,
  ``D1-brane in beta-Deformed Background,''
  JHEP {\bf 0711}, 011 (2007)
  [arXiv:0710.0148 [hep-th]].

\bibitem{Hofman:2006xt}
  D.~M.~Hofman and J.~M.~Maldacena,
``Giant magnons,''
  J.\ Phys.\ A  {\bf 39}, 13095 (2006)
  [arXiv:hep-th/0604135].

\bibitem{Ishizeki:2007we}
  R.~Ishizeki and M.~Kruczenski,
  ``Single spike solutions for strings on S**2 and S**3,''
  Phys.\ Rev.\ D {\bf 76}, 126006 (2007)
  doi:10.1103/PhysRevD.76.126006
  [arXiv:0705.2429 [hep-th]].

\bibitem{Stepanchuk:2012xi}
  A.~Stepanchuk and A.~A.~Tseytlin,
  ``On (non)integrability of classical strings in p-brane backgrounds,''
  J.\ Phys.\ A {\bf 46}, 125401 (2013)
  [arXiv:1211.3727 [hep-th]].



\bibitem{Kluson:2006wa}
  J.~Kluson,
  ``Dynamics of probe brane in the background of intersecting fivebranes,''
  Phys.\ Rev.\ D {\bf 73}, 106008 (2006)
  [hep-th/0601229].

  \bibitem{Kluson:2007st}
  J.~Kluson,
  ``Fundamental string and D1-brane in I-brane background,''
  JHEP {\bf 0801}, 045 (2008)
  [arXiv:0711.4219 [hep-th]].

  \bibitem{Itzhaki:2005tu}
  N.~Itzhaki, D.~Kutasov and N.~Seiberg,
  ``I-brane dynamics,''
  JHEP {\bf 0601}, 119 (2006)
  [hep-th/0508025].


 \bibitem{Cagnazzo:2012se}
A.~Cagnazzo and K.~Zarembo, {\it {B-field in {$AdS_3/CFT_2$} Correspondence and
  Integrability}},  {\em JHEP} {\bf 1211} (2012) 133,
  [\href{http://xxx.lanl.gov/abs/1209.4049}{{\tt arXiv:1209.4049}}].

\bibitem{Wulff:2014kja}
L.~Wulff, {\it {Superisometries and integrability of superstrings}},
  \href{http://xxx.lanl.gov/abs/1402.3122}{{\tt arXiv:1402.3122}}.

  \bibitem{Hoare:2013pma}
B.~Hoare and A.~Tseytlin, {\it {On string theory on {$AdS_3 \times S^3\times
  T^4$} with mixed 3-form flux: Tree-level S-matrix}},  {\em Nucl.Phys.} {\bf
  B873} (2013) 682--727, [\href{http://xxx.lanl.gov/abs/1303.1037}{{\tt
  arXiv:1303.1037}}].

\bibitem{Hoare:2013ida}
B.~Hoare and A.~Tseytlin, {\it {Massive S-matrix of $AdS_3 \times S^3 \times
  T^4$ superstring theory with mixed 3-form flux}},  {\em Nucl.Phys.} {\bf
  B873} (2013) 395--418, [\href{http://xxx.lanl.gov/abs/1304.4099}{{\tt
  arXiv:1304.4099}}].

\bibitem{Bianchi:2014rfa}
L.~Bianchi and B.~Hoare, {\it {$AdS_3 \times S^3 \times M^4$ string S-matrices
  from unitarity cuts}},  \href{http://xxx.lanl.gov/abs/1405.7947}{{\tt
  arXiv:1405.7947}}.

\bibitem{Hoare:2013lja}
B.~Hoare, A.~Stepanchuk, and A.~Tseytlin, {\it {Giant magnon solution and
  dispersion relation in string theory in {$AdS_3\times S^3\times T^4$} with
  mixed flux}},  {\em Nucl.Phys.} {\bf B879} (2014) 318--347,
  [\href{http://xxx.lanl.gov/abs/1311.1794}{{\tt arXiv:1311.1794}}].

\bibitem{Babichenko:2014yaa}
A.~Babichenko, A.~Dekel, and O.~Ohlsson~Sax, {\it {Finite-gap equations for
  strings on $AdS_3 x S^3 x T^4$ with mixed 3-form flux}},
  \href{http://xxx.lanl.gov/abs/1405.6087}{{\tt arXiv:1405.6087}}.

\bibitem{Borsato:2014hja}
  R.~Borsato, O.~Ohlsson Sax, A.~Sfondrini and B.~Stefanski,
  ``The complete AdS$_{3} \times$ S$^3 \times$ T$^4$ worldsheet S matrix,''
  JHEP {\bf 1410}, 66 (2014)
  [arXiv:1406.0453 [hep-th]].

\bibitem{Lloyd:2014bsa}
  T.~Lloyd, O.~Ohlsson Sax, A.~Sfondrini and B.~Stefanski, Jr.,
  ``The complete worldsheet S matrix of superstrings on $AdS_3 x S^3 x T^4$ with mixed three-form flux,''
  Nucl.\ Phys.\ B {\bf 891}, 570 (2015)
  [arXiv:1410.0866 [hep-th]].

  \bibitem{David:2014qta}
  J.~R.~David and A.~Sadhukhan,
  ``Spinning strings and minimal surfaces in $AdS_3$ with mixed 3-form fluxes,''
  JHEP {\bf 1410}, 49 (2014)
  [arXiv:1405.2687 [hep-th]].

\bibitem{Ahn:2014tua}
  C.~Ahn and P.~Bozhilov,
  ``String solutions in $AdS_3 x S^3 x T^4$ with NS-NS B-field,''
  Phys.\ Rev.\ D {\bf 90}, no. 6, 066010 (2014)
  [arXiv:1404.7644 [hep-th]].

\bibitem{misc9}
  A.~Banerjee, K.~L.~Panigrahi and P.~M.~Pradhan,
  ``Spiky strings on $AdS_3 \times S^3$ with NS-NS flux,''
  Phys.\ Rev.\ D {\bf 90}, no. 10, 106006 (2014)
  [arXiv:1405.5497 [hep-th]].

\bibitem{Hernandez:2014eta}
  R.~Hernández and J.~M.~Nieto,
  ``Spinning strings in $AdS_3 \times S^3$ with NS–NS flux,''
  Nucl.\ Phys.\ B {\bf 888}, 236 (2014)
  [Nucl.\ Phys.\ B {\bf 895}, 303 (2015)]
  [arXiv:1407.7475 [hep-th]].

\bibitem{Hernandez:2015nba}
  R.~Hernandez and J.~M.~Nieto,
  ``Elliptic solutions in the Neumann–Rosochatius system with mixed flux,''
  Phys.\ Rev.\ D {\bf 91}, no. 12, 126006 (2015)
  [arXiv:1502.05203 [hep-th]].

\bibitem{Banerjee:2015bia}
  A.~Banerjee, K.~L.~Panigrahi and M.~Samal,
  ``A note on oscillating strings in AdS$_{3}$ × S$^{3}$ with mixed three-form fluxes,''
  JHEP {\bf 1511}, 133 (2015)
  [arXiv:1508.03430 [hep-th]].

\bibitem{Banerjee:2015qeq}
  A.~Banerjee and A.~Sadhukhan,
  ``Multi-spike strings in $AdS_3$ with mixed three-form fluxes,''
  arXiv:1512.01816 [hep-th].

\bibitem{Kluson:2015lia}
  J.~Kluson,
  ``Integrability of D1-brane on Group Manifold with Mixed Three Form Flux,''
  arXiv:1509.09061 [hep-th].

  \bibitem{Biswas:2011wu}
  S.~Biswas and K.~L.~Panigrahi,
  ``Spiky Strings on NS5-branes,''
  Phys.\ Lett.\ B {\bf 701}, 481 (2011)
  [arXiv:1103.6153 [hep-th]].

\bibitem{Biswas:2012wu}
  S.~Biswas and K.~L.~Panigrahi,
  ``Spiky Strings on I-brane,''
  JHEP {\bf 1208}, 044 (2012)
  [arXiv:1206.2539 [hep-th]].

\bibitem{Burgess:2003mm}
  C.~P.~Burgess, N.~E.~Grandi, F.~Quevedo and R.~Rabadan,
  ``D-brane chemistry,''
  JHEP {\bf 0401}, 067 (2004)
  [hep-th/0310010].
  
\bibitem{Chen:2008vc} 
  C.~M.~Chen, J.~H.~Tsai and W.~Y.~Wen,
  ``Giant Magnons and Spiky Strings on $S^3$ with B-field,''
  Prog.\ Theor.\ Phys.\  {\bf 121}, 1189 (2009)
  doi:10.1143/PTP.121.1189
  [arXiv:0809.3269 [hep-th]].
\end{thebibliography}
\end{document}